%% file: main.tex
\documentclass[a4paper]{JHEP3}

\usepackage{graphicx}
\usepackage{amsmath}
\usepackage{cite}
\usepackage{color}
\usepackage{listings}
\let\normalcolor\relax

\usepackage{multirow}
\usepackage{mathenv}
\usepackage{array}

\definecolor{shaded}{RGB}{210,210,210}
\def\eqref#1{(\ref{#1})}
\def\text{\rm }
\def\Res{{\rm Res}}
\def\samurai{{{\sc samurai}}}
\def\oo{\infty}

\newcommand{\beq}{\begin{equation}}
\newcommand{\eeq}{\end{equation}}
\newcommand{\bqa}{\begin{eqnarray}}
\newcommand{\eqa}{\end{eqnarray}}

\def\db#1{ D_{#1}}

\def\slh#1{\rlap / {#1}}

\def\Formcalc{{{\sc FormCalc}}}
\def\Gosam{{{\sc GoSam}}}
\def\mathematica{{{\sc mathematica}}}
\def\sam{{{\sc S@M }}}
\def\form{{{\sc form }}}

\def\samurai{{{\sc samurai }}}
\def\cuttools{{{\sc CutTools }}}
\def\C++{{{\sc c++ }}}

\def\ra{\rangle}

\def\spa#1.#2{\langle#1\,#2\rangle}
\def\spb#1.#2{[#1\,#2]}

\def\spab#1.#2.#3{\langle\mskip-1mu{#1}
                  | #2 | {#3}]}

\def\spba#1.#2.#3{[\mskip-1mu{#1}
                  | #2 | {#3}\rangle}

\def\spbb#1.#2.#3.#4{[\mskip-1mu{#1}
                     | {#2} \ {#3} | {#4}]}

\def\spaa#1.#2.#3.#4{\langle\mskip-1mu{#1}
                     | {#2} \ {#3} | {#4}\rangle}

\newcommand{\bea}{\begin{eqnarray}}

\newcommand{\eea}{\end{eqnarray}}

\newcommand{\bean}{\begin{eqnarray*}}

\newcommand{\eean}{\end{eqnarray*}}
           
\newcommand{\nn}{\nonumber \\}

\title{Integrand reduction of one-loop scattering  amplitudes through Laurent series expansion
}

\author{Pierpaolo  Mastrolia \\ 
Max-Planck-Institut f\"ur Physik, F\"ohringer Ring 6, D-80805 M\"unchen, Germany; \\
Dipartimento di Fisica e Astronomia, Universit\`a di Padova, and INFN Sezione di Padova, via Marzolo 8, 35131 Padova, Italy \\
E-mail: \email{pierpaolo.mastrolia@cern.ch}
}    

\author{Edoardo Mirabella \\
Max-Planck-Institut f\"ur Physik, F\"ohringer Ring 6, D-80805 M\"unchen, Germany \\
E-mail: \email{mirabell@mppmu.mpg.de}
}

\author{Tiziano Peraro \\
Max-Planck-Institut f\"ur Physik, F\"ohringer Ring 6, D-80805 M\"unchen, Germany \\
E-mail: \email{peraro@mppmu.mpg.de}
}        

\preprint{MPI-2012-51}

\abstract{
We present a semi-analytic method for the integrand reduction of one-loop amplitudes, based on the  
systematic application of the Laurent expansions to the integrand-decomposition. In the 
asymptotic limit, the coefficients of the master integrals  are the  solutions of a 
diagonal system of equations, properly corrected by counterterms whose parametric form is known a 
priori. 
The Laurent expansion of the integrand is implemented through  polynomial division. 
The extension of the integrand-reduction to the case of numerators with rank larger than the number 
of propagators is discussed as well.
}

\begin{document}

\input{introduction}

\input{reductions}

\input{method}

\input{implementation}

\input{example}

\input{extension}

\input{conclusions}

\section*{Acknowledgments}
We would like to thank Simon Badger, Gudrun Heinrich, Giovanni Ossola and Thomas Reiter for interesting 
discussions and feedback on the manuscript.  This work was 
supported by the Alexander von Humboldt Foundation, in the framework of the Sofja Kovaleskaja
Award, endowed by the German Federal Ministry of Education and Research.

\appendix

\input{AppendixN}

\input{AppendixM}

\bibliographystyle{JHEP}

\bibliography{references}

\end{document}

%% file: introduction.tex
\section{Introduction}

The recent development of novel methods for computing one-loop scattering amplitudes
has been highly stimulated by a deeper understanding 
of their multi-channel factorization properties under
special kinematics enforced by on-shellness \cite{Cachazo:2004kj,Britto:2004ap} and  
generalized unitarity \cite{Bern:1994zx,Britto:2004nc}.
Analyticity and unitarity of scattering amplitudes 
have then been strengthened  by the complementary classification of the mathematical 
structures present in the residues at the singular points. They naturally arise  
after uncovering a relation between numerator and denominators 
of one-loop Feynman integr{\it als}, yielding the multipole decomposition of 
Feynman integr{\it ands} originally proposed in a four-dimensional framework
by Ossola, Papadopoulos and Pittau (OPP) \cite{Ossola:2006us,Ossola:2007bb}, and later 
extended to dimensionally regulated amplitudes by Ellis, Giele, Kunszt and 
Melnikov (EGKM) \cite{Ellis:2007br,Giele:2008ve,Ellis:2008ir}.

The use of unitarity-cuts and complex momenta for on-shell internal particles 
turned unitarity-based methods into very efficient
tools for computing scattering amplitudes. 
These methods, recently reviewed in \cite{Berger:2009zb,Britto:2010xq,Ellis:2011cr}, exploit two general properties of 
scattering amplitudes, such as analyticity and unitarity:
the former granting that amplitudes can be reconstructed from 
the knowledge of their (generalized) singularity-structure; the latter granting that 
the residues at the singular points factorize into products of simpler amplitudes.
Unitarity-based methods are founded on the underlying representation 
of scattering amplitudes as a linear combination of master integrals (MI's)~\cite{Passarino:1978jh,vanOldenborgh:1989wn}, 
and their principle is the extraction of the coefficients entering in such a linear combination 
by matching the cuts of the amplitudes onto the cuts of each MI.

The multi-particle pole decomposition for the integrands 
of arbitrary scattering amplitudes emerges
from the combination of analyticity and unitarity 
with the idea of a reduction under the integral sign. 

The principle of an integrand-reduction method is 
the underlying multi-particle pole expansion for the integrand of 
any scattering amplitude, or, equivalently, 
a representation where the numerator of each Feynman integral is expressed as 
a combination 
of products of the corresponding denominators, with polynomial coefficients.
Each residue is a (multivariate) polynomial in the 
{\it irreducible scalar products} (ISP's) formed by the loop momenta and 
either external momenta or polarization vectors constructed out of them.
The {\it integrand reduction method} has been recently shown to be applicable to scattering
amplitudes beyond one-loop as well~\cite{Mastrolia:2011pr, Badger:2012dp}.

The polynomial structure of the multi-particle residues is a {\it qualitative} information
that turns into a {\it quantitative} algorithm for decomposing arbitrary 
amplitudes in terms of MI's at the integrand level. 
In the context of an integrand-reduction, any explicit integration procedure 
and/or any matching procedure between cuts of amplitudes and cuts of MI's 
is replaced by {\it polynomial fitting}, which is a simpler operation.
Within this algorithm the (dimensionally regulated) integrand of a given scattering amplitude
is the only input needed for  sampling  the integrand on the solutions 
of generalized on-shell conditions.

The original algorithm \cite{Ossola:2006us,Ellis:2007br} is based on the solution 
of a triangular system of equations to be solved {\it top-down}, from the 
determination of the 5-point coefficients to the 1-point coefficients. 
At any step of the reduction, the  Gauss-substitutions 
requires the subtraction of the  of the coefficients determined in the previous steps.
Accordingly, the algorithm  proceeds by {\it subtractions at the integrand level},
requiring the knowledge of the already reconstructed residues. In other words
the determination of the coefficients of the $n$-point residues
$(1 \le n \le 5)$, 
requires the subtraction of the residues of $m$-point functions with $n< m \le 5$.
The integrand reduction method has been implemented in the publicly available 
libraries {\cuttools}~\cite{Ossola:2007ax} and {\samurai}~\cite{Mastrolia:2010nb}, as
well as in several multi-purpose 
codes~\cite{Berger:2008sj,Giele:2008bc,Lazopoulos:2008ex,Winter:2009kd,vanHameren:2009dr,Bevilacqua:2010mx,Badger:2010nx,Bevilacqua:2011xh,Hirschi:2011pa,Cullen:2011ac,Agrawal:2011tm,Hahn:2010zi}.

In this paper, we exploit the asymptotic behaviour of the integrand decomposition, 
and propose a simpler technique for the {\it integrand-reduction} of one-loop scattering 
amplitudes in dimensional regularization.
The use of {\it Laurent series} within the unitarity-based methods has been mainly 
developed in the context of analytic calculations 
\cite{Britto:2007tt,Britto:2008vq,Forde:2007mi,Kilgore:2007qr,Badger:2008cm}.

Elaborating on the the techniques proposed by Forde \cite{Forde:2007mi} and 
Badger \cite{Badger:2008cm} we apply systematically the series expansion to 
the integrand decomposition formula of OPP/EGKM. We show how the advantages of the 
analytic techniques can be incorporated in a refined semi-analytic algorithm 
which determines the  coefficients  using 
Laurent expansions  rather than  polynomial fitting.
The main features of this algorithm are the following.
\begin{itemize}
\item The coefficients of 5-point functions are never needed and do not
      have to be computed.
\item The   spurious coefficients  of the 4-point functions do not enter the reduction and are not computed.
      The  rational terms coming from higher-dimension 4-point functions can be 
      computed analytically  from quadruple-cuts only.
\item The computation of 3-, 2-, and 1-point coefficients is independent of the 
      residues of the 4-point functions. In particular the 3-point coefficients
      are computed from triple cuts, without any subtraction.
\item  The  subtraction at the integrand level is replaced by  the 
       subtraction at the {\it coefficient level}. Indeed in the original 
       reduction method the subtractions guarantee the polynomiality of each residue  
       allowing its determination through polynomial 
       fitting. The Laurent expansion makes each function entering the subtraction separately 
       polynomial. Therefore the subtraction of higher point residues can be omitted
       during the reduction. Its effect is accounted for correcting the reconstructed
       coefficients.
\item  The correction terms of  2-, and 1-point functions are parametrized 
       by universal functions in terms of the higher points  coefficients.
\item The application of the Laurent expansion for the determination of 3-, 2-, 
      and 1-point coefficients is implemented via {\it polynomial division}. 
      The Laurent series is obtained as the {\it quotient} of the division between 
      the numerator and the product of the uncut denominators, {\it neglecting the remainder}.
\end{itemize}

This algorithm has been implemented in {\C++} and in {\mathematica} using the {\sam} 
package \cite{Maitre:2007jq}. The semi-analytic implementation in {\C++} has been 
designed as a reduction library to be linked
to codes like {\Gosam}  and {\Formcalc} which provide analytic expression of the integrands,
or to any package which can provide the tensor-structure of the integrand 
as described in \cite{vanHameren:2009vq,Heinrich:2010ax} and in the more recent 
Open-Loop technique \cite{Cascioli:2011va}. The version in {\mathematica} has been used to obtain 
closed formulas for the coefficients, depending on the vector-basis associated to each cut and 
on the generic tensors appearing in the integrand. 
In this latter fashion, the reduction procedure is replaced by a simple pattern-matching.

%% file: reductions.tex
\section{Integrand decomposition}
\label{Sec:IntegrandR}

In this section we collect the relevant formulae of the integrand reduction
methods introduced in~\cite{Ossola:2006us,Ossola:2007bb,Ellis:2007br,Giele:2008ve,Ellis:2008ir} following 
the notation of~\cite{Mastrolia:2010nb}.

The reduction method is based on the general 
decomposition for the integrand  of a generic  
one-loop amplitude, Any one-loop $n$-point amplitude 
can be written as
\bea
&& {\cal A}_n = \int d^d {\bar q} \ A(\bar q, \epsilon) \ , \nn 
&& A(\bar q, \epsilon)= \frac{{\cal N}({\bar q}, \epsilon)}{\db{0}\db{1}\cdots \db{n-1}} \ , \nn
&& \db{i} = ({\bar q} + p_i)^2-m_i^2 = 
            (q + p_i)^2-m_i^2-\mu^2, \qquad (p_0 \ne 0)\,.
\label{def:An}
\eea
We use a bar to denote objects living in $d=~4-2\epsilon$  
dimensions, following the prescription
\bea \label{eq:qmu}
 \slh{{\bar q}} = \slh{q} + \slh{\mu} \ , \quad {\rm with} \qquad 
{\bar q}^2= q^2 - \mu^2 \ . 
\eea
The most general numerator of one-loop amplitudes
$\mathcal{N}(\bar{q}, \epsilon)$ can be thought as composed of three terms,
\begin{equation}
\mathcal{N}(\bar{q}, \epsilon) =
            N_0(q, \mu^2)
+ \epsilon  N_1(q, \mu^2)
+\epsilon^2 N_2(q, \mu^2).
\label{eq:genericN}
\end{equation}
The coefficients of this $\epsilon$-expansion,  $N_0$, $N_1$ and $N_2$, are functions of
$q^\nu$ and $\mu^2$.  In the following  discussion we will 
denote by $N$ any element of the set $\{ N_0, N_1, N_2\}$.

The numerator $N(q,\mu^2)$ can be expressed in terms of denominators $\db{i}$, as follows
\bea
\label{eq:2}
N(q,\mu^2) &=&
\sum_{i < \!< m}^{n-1}
          \Delta_{ i j k \ell m}(q,\mu^2)
\prod_{h \ne i, j, k, \ell, m}^{n-1} \db{h} 
+\sum_{i < \!< \ell}^{n-1}
          \Delta_{ i j k \ell }(q,\mu^2)
\prod_{h \ne i, j, k, \ell}^{n-1} \db{h} 
+    \pagebreak[1]  \nn     &+&
\sum_{i < \!< k}^{n-1}
          \Delta_{i j k}(q,\mu^2)
\prod_{h \ne i, j, k}^{n-1} \db{h} 
+\sum_{i < j }^{n-1}
          \Delta_{i j}(q,\mu^2) 
\prod_{h \ne i, j}^{n-1} \db{h} 
+\sum_{i}^{n-1}
          \Delta_{i}(q,\mu^2) 
\prod_{h \ne i}^{n-1} \db{h} \ , \qquad
\label{def:OPP:deco}
\eea
where $ i < \!< m $ is the  lexicographic ordering $i < j < k < \ell < m$.
The functions $\Delta(q,\mu^2)$ are polynomials in the components 
of $q$ and in $\mu^2$. The decomposition (\ref{def:OPP:deco}) 
expose the multi-pole nature of the integrand
\bea
A(q,\mu^2) &=&
\sum_{i < \!< m}^{n-1}
         { \Delta_{ i j k \ell m}(q,\mu^2) \over 
           \db{i} \db{j} \db{k} \db{\ell} \db{m} } 
+
\sum_{i < \!< \ell}^{n-1}
         { \Delta_{ i j k \ell }(q,\mu^2) \over 
           \db{i} \db{j} \db{k} \db{\ell} } 
+
\sum_{i < \!<  k }^{n-1}
         { \Delta_{i j k}(q,\mu^2) \over 
           \db{i} \db{j} \db{k} }
+  \pagebreak[1]  \nn &+& 
\sum_{i < j }^{n-1}
         { \Delta_{i j}(q,\mu^2) \over 
           \db{i} \db{j} } 
+
\sum_{i}^{n-1}
         { \Delta_{i}(q,\mu^2) \over
           \db{i} } \ .
\label{def:OPP:aaa}
\eea

For each cut  $(ijk\cdots)$,  obtained setting $D_i=D_j=D_k=\cdots=0$, we  
introduce a basis of four massless
vectors
\bea
\mathcal{E}^{(ijk\cdots)} = \left \{ e^{(ijk\cdots)}_1, e^{(ijk\cdots)}_2, e^{(ijk\cdots)}_3, e^{(ijk\cdots)}_4 \right   \} \ ,
\label{Eq:Eps}
\eea
such that
\begin{align*}
 \left (e^{(ijk\cdots)}_i \right )^2 = 0 \ , &\qquad             e^{(ijk\cdots)}_1 \cdot e^{(ijk\cdots)}_3 = e^{(ijk\cdots)}_1 \cdot e^{(ijk\cdots)}_4= 0\ ,  \nonumber \\
 e^{(ijk\cdots)}_2 \cdot e^{(ijk\cdots)}_3 = e^{(ijk\cdots)}_2 \cdot e^{(ijk\cdots)}_4 = 0 \ , &\qquad
 e^{(ijk\cdots)}_1 \cdot e^{(ijk\cdots)}_2 = - e^{(ijk\cdots)}_3 \cdot e^{(ijk\cdots)}_4 = 1 \ . \nonumber 
\end{align*}
The massless vectors $e^{(ijk\cdots)}_1$ and $e^{(ijk\cdots)}_2$ can be written as a linear combination of the two external
legs at the edges of the propagator carrying momentum $q+p_i$, say $K_1$ and $K_2$, along the lines 
of~\cite{Mastrolia:2010nb}.  In the 
case  of 
double-cut, $K_1$ is the momentum flowing through the  corresponding 2-point diagram, and $K_2$ is an arbitrary massless vector. In 
the case of single-cut both  $K_1$ and $K_2$  are chosen as arbitrary vectors. In the case of  quadruple-cut $(ijk\ell)$
we define
\bea
v^{(ijk\ell)}_\perp &=& \left (K_3 \cdot  e^{(ijk\ell)}_4 \right ) \ e^{(ijk\ell)}_3 - \left (K_3 \cdot  e^{(ijk\ell)}_3 \right ) \ e^{(ijk\ell)}_4, \nonumber \\
v^{(ijk\ell)} &=&  \left (K_3 \cdot  e^{(ijk\ell)}_4 \right ) \ e^{(ijk\ell)}_3 + \left (K_3 \cdot  e^{(ijk\ell)}_3 \right ) \ e^{(ijk\ell)}_4 \  .
\label{Eq:VVp}
\eea
The momentum $K_3$ is the third leg of the 4-point function associated to the considered quadruple-cut. To simplify our  notation we will
omit the indices of the cut $(ijk\cdots)$ whenever possible.

The functions $\Delta(q,\mu^2)$ are parametrized in terms of the 
basis~(\ref{Eq:Eps}) and of the vectors~(\ref{Eq:VVp}):
\bea
\label{def:Resi5:right}
\Delta_{i j k \ell m}(q,\mu^2) &=& c_{5,0}^{( i j k \ell m)} \ \mu^2 \ , \\[1.0ex]
\label{def:Resi4:right}
 \Delta_{ i j k \ell}(q,\mu^2) &=& \Delta_{ i j k \ell}^{R}(q,\mu^2) + 
 c_{4,0}^{( i j k \ell)}
+c_{4,2}^{( i j k \ell)} \mu^2
+c_{4,4}^{( i j k \ell)} \mu^4  \ ,    \pagebreak[1] \\[1.0ex]
\label{def:Resi3:right}
\Delta_{ i j k}(q,\mu^2) 
&=& 
 \Delta_{ i j k }^{R}(q,\mu^2) +
 c_{3,0}^{( i j k)} +c_{3,7}^{( i j k)} \mu^2  \ ,  \pagebreak[1]  \\[1.0ex]
\label{def:Resi2:right}
\Delta_{ i j}(q,\mu^2) 
&=& 
 \Delta_{ i j  }^{R}(q,\mu^2) +
 c_{2,0}^{(i j)} +c_{2,9}^{(i j)} \mu^2   \ ,  \pagebreak[1]   \\[1.0ex]
\label{def:Resi1:right}
\Delta_{ i }(q,\mu^2) 
&=& 
  c_{1,0}^{(i)} 
+ c_{1,1}^{(i)} ((q+p_i)\cdot e_1)
+ c_{1,2}^{(i)} ((q+p_i)\cdot e_2)   \pagebreak[1]  \nn
&+& c_{1,3}^{(i)} ((q+p_i)\cdot e_3)
+ c_{1,4}^{(i)} ((q+p_i)\cdot e_4)  \ .
\eea 
For later convenience, we define the {\it reduced} polynomials $\Delta^{R}$ as,
\bea
\label{def:Resi4R:right}
 \Delta_{ i j k \ell}^{R}(q,\mu^2) &=&
\Big(
c_{4,1}^{( i j k \ell)}
+c_{4,3}^{( i j k \ell)} \ \mu^2
\Big)  
 (q+p_i)\cdot v_\perp \ ,    \pagebreak[1] \\[1.0ex]
\label{def:Resi3R:right}
\Delta_{ i j k}^{R}(q,\mu^2) 
&=& 
 \Big(c_{3,1}^{( i j k)} + c_{3,8}^{( i j k)} \mu^2\Big) (q+p_i)\cdot e_3  + \Big(c_{3,4}^{( i j k)} + c_{3,9}^{( i j k)} \mu^2\Big)(q+p_i)\cdot e_4
\pagebreak[1] \nn
&+&c_{3,2}^{( i j k)} ((q+p_i)\cdot e_3)^2  +c_{3,5}^{( i j k)} ((q+p_i)\cdot e_4)^2  \pagebreak[1]  \nn
&+&c_{3,3}^{( i j k)} ((q+p_i)\cdot e_3)^3  +c_{3,6}^{( i j k)} ((q+p_i)\cdot e_4)^3 \ , \pagebreak[1]   \\[1.0ex]
\label{def:Resi2R:right}
\Delta_{ i j}^{R}(q,\mu^2) 
&=&
c_{2,1}^{(i j)} (q+p_i)\cdot e_2
+c_{2,2}^{(i j)} ((q+p_i)\cdot e_2)^2   \pagebreak[1]  \nn
&+&c_{2,3}^{(i j)} (q+p_i)\cdot e_3
+c_{2,4}^{(i j)} ((q+p_i)\cdot e_3)^2   \pagebreak[1]  \nn
&+&c_{2,5}^{(i j)} (q+p_i)\cdot e_4
+c_{2,6}^{(i j)} ((q+p_i)\cdot e_4)^2   \pagebreak[1]   \nn
&+&c_{2,7}^{(i j)} ((q+p_i)\cdot e_2) ((q+p_i)\cdot e_3)
+c_{2,8}^{(i j)} ((q+p_i)\cdot e_2) ((q+p_i)\cdot e_4) \, . \
\eea

Neglecting terms of $\mathcal{O}(\epsilon)$, the one loop amplitude can be written in terms of  master integrals and of the coefficients of 
$\Delta_{ i j k \ell m}$, $\Delta_{ i j k \ell}$, $\Delta_{ i j k}$, $\Delta_{ i j}$,  
and $\Delta_{ i }$,
\bea
{\cal A}_n &=& 
\sum_{i < j < k < \ell}^{n-1}\bigg\{
          c_{4,0}^{ (i j k \ell)} I_{i j k \ell} +  
          c_{4,4}^{ (i j k \ell)} I_{i j k \ell}[\mu^4] 
\bigg\}   \pagebreak[0]  \nn
     &+&
\sum_{i < j < k }^{n-1}\bigg\{
          c_{3,0}^{ (i j k)} I_{i j k} +
          c_{3,7}^{ (i j k)} I_{i j k}[\mu^2]
\bigg\}  \pagebreak[0]  \nn
     &+&
\sum_{i < j }^{n-1}\bigg\{
          c_{2,0}^{ (i j)} I_{i j} 
        + c_{2,1}^{ (i j)} I_{i j}[(q+p_i)\cdot e_2 ] 
        + c_{2,2}^{ (i j)} I_{i j}[((q+p_i)\cdot e_2)^2 ]         +  c_{2,9}^{ (i j)} I_{i j}[\mu^2]  \bigg\}  \pagebreak[0]  \nn 
&+& 
\sum_{i}^{n-1}
      c_{1,0}^{ (i)} I_{i} 
 \ ,
\label{eq:Aresult}
\eea
where
\bea
 I_{i_1 \cdots i_k}[\alpha] \equiv \int d^d \bar q {\alpha \over \db{i_1} \cdots \db{i_k} }, 
\qquad
I_{i_1 \cdots i_k} \equiv   I_{i_1 \cdots i_k}[1].
\eea
As already noted in~\cite{Ossola:2006us,Giele:2008ve,Ellis:2011cr}, 
some of the terms appearing in the integrand decomposition~(\ref{def:OPP:aaa}) vanish
upon integration. They are called   {\it spurious}  and  do not contribute 
to the amplitude~(\ref{eq:Aresult}).  Beside the scalar boxes, triangles, bubbles and tadpoles, 
the other master integrals are the linear and quadratic two-points 
functions\cite{Pittau:1996ez,Bern:1995db} and the integrals containing 
powers of $\mu^2$ in the numerator. The latter 
can be traded with higher dimensional
integrals~\cite{Pittau:1996ez,Bern:1995db}
\bea
 I_{i_1 \cdots i_k}[(\mu^2)^r f(q, \mu^2)]  =  \frac{1}{\pi^r} \prod_{\kappa=1}^{r}  \left ( \kappa -3 +   { d \over 2 } \right )
  \int d^{d+2r} \bar q { f(q, \mu^2 ) \over \db{i_1} \cdots \db{i_k} } \ .  
\eea
As already noticed in \cite{Melnikov:2010iu}, eq. (\ref{eq:Aresult})  is free of scalar pentagons.

%% file: method.tex
\section{Reduction algorithm}
\label{sec:Reduction}
\label{SSec:MethodCoeff}
The one loop amplitude is completely known, provided the coefficients $c$ appearing the 
r.h.s. of eq.~(\ref{eq:Aresult})  are known.  In  this subsection we show  how to get 
the coefficients of each polynomial $\Delta$ performing suitable  series expansions.

\subsection*{Quintuple Cut}

The coefficient  $ c_{5,0}^{( i j k \ell m)}$, eq.~(\ref{def:Resi5:right}),
can be  computed using a quintuple cut. However its actual value it is not relevant in 
our reduction algorithm. Therefore its computation is omitted.

\subsection*{Quadruple cut}
The solutions of the quadruple cut $\db{i} = \ldots = \db{\ell} = 0$, can be expressed as
\bea
q^{(ijk\ell)}_\pm 
=  - p_i + x_1 e_1 + x_2 e_2 
+ x_v v \pm  u \;   v_\perp \, \qquad u=  \sqrt{\alpha_\perp + \frac{\mu^2}{v^2_\perp}  }
\eea
where the coefficients  $x_1, x_2, x_v$ and $\alpha_\perp$ are fully determined by the 
cut-conditions. The only two coefficients that are needed are obtained 
from~\cite{Britto:2004nc, Badger:2008cm}.
\bea
&& {1 \over 2} \left [ 
{N(q^{(ijk\ell)}_+ ,0) \over   \prod_{h\neq i,j,k,\ell} \db{h}(q^{(ijk\ell)}_+ , 0) } +
{N(q^{(ijk\ell)}_- ,0) \over   \prod_{h\neq i,j,k,\ell} \db{h}(q^{(ijk\ell)}_- , 0) } \right ]
=    c^{(ijk\ell)}_{4,0} \ ,  \\
&& {N_\pm \over   \prod_{h\neq i,j,k,\ell} \db{h , \pm} } \Bigg |_{\mu^2 \to \infty}  =
  c_{4,4}^{(ijk\ell)} \mu^4 + \mathcal{O}(\mu^3) \ .
\label{Eq:Quad2}
\eea 
Here and in  the following we  use the abbreviation
\bea
f_{\pm} \equiv f \left (q^{(ijk\ell)}_{\pm} , \mu^2 \right ),
\eea
omitting  whenever possible the indices of the cuts $\{i,j, \ldots  \} $ as well as the $\mu^2$ dependence.

\subsection*{Triple cut}
The solutions of the  triple-cut, 
$\db{i} = \db{j} = \db{k} = 0$ can be parametrized as,
\bea
q^{(ijk)}_+ &=& - p_i + x_1 e_1 + x_2 e_2 + t e_3 + {\alpha_0 + \mu^2 \over t} e_4 \ , \\
q^{(ijk)}_- &=& - p_i + x_1 e_1 + x_2 e_2 + {\alpha_0 + \mu^2 \over t } e_3 + t e_4 \ , 
\eea
where $x_1, x_2$ and $\alpha_0$ are frozen by the triple-cut conditions.  
The coefficients can be obtained in the large-$t$ limit according to
\bea
\label{Eq:Triple1}
&&  {N_{+} \over \prod_{h\neq i,j,k} \db{h, +}}  \Bigg |_{t \to \infty}  = - \Big(c_{3,4}^{( i j k)} + c_{3,9}^{( i j k)} \mu^2\Big)  \ t  + 
 c_{3,5}^{( i j k)}  \ t^2 - c_{3,6}^{( i j k)}  \ t^3   + \mathcal{O}(1) \ , \\
 \label{Eq:Triple2}
&&  {N_{-} \over \prod_{h\neq i,j,k} \db{h, -}} \Bigg |_{t \to \infty}  = 
- \Big(c_{3,1}^{( i j k)} + c_{3,8}^{( i j k)} \mu^2\Big) \ t  + 
 c_{3,2}^{( i j k)} \ t^2 - c_{3,3}^{( i j k)}  \ t^3 + \mathcal{O}(1) \ , \\
&&{1 \over 2} \left [ {N_{+} \over \prod_{h\neq i,j,k} \db{h, +}} +  {N_{-} \over \prod_{h\neq i,j,k} \db{h, -}} \right ] \Bigg |_{t \to \infty} =  
 \Big(c_{3,0}^{( i j k)} + c_{3,7}^{( i j k)} \mu^2\Big) + \mathcal{O}\left ( {1 \over t}  \right) +  
\Omega(t) \, . \ \
\label{Eq:Triple3}
\eea
The Bachmann-Landau symbol $\Omega(t)$ denotes terms which are non-negligible with respect to $t$ as $t \to \oo$.
Eq.~(\ref{Eq:Triple3}) has been introduced in~\cite{Forde:2007mi,Badger:2008cm}. 
As explained in~\cite{Ellis:2011cr},
the average over the two solutions cancel the contributions of the spurious coefficient of the boxes.
Eqs.~(\ref{Eq:Triple1}) and~(\ref{Eq:Triple2}) determine the spurious coefficients of the 
triangles. They are independent of the  spurious coefficients of the boxes which  are
$\mathcal{O}(1)$ in the $t \to \oo$ expansion.

\subsection*{Double cut}
The solutions of the double cut $\db{i} = \db{j} = 0$  are parametrized as follows
\bea
q^{(ij)}_+ &=& - p_i + x_1 e_1 + (\alpha_0 + x_1 \alpha_1) e_2 + t e_3 + { \mu^2 + \beta_0 + \beta_1 x_1 + \beta_2 x_1^2 \over 2 \  t} e_4 \ , \nonumber \\
q^{(ij)}_- &=& - p_i + x_1 e_1 + (\alpha_0 + x_1 \alpha_1) e_2 + {  \mu^2 + \beta_0 + \beta_1 x_1 + \beta_2 x_1^2 \over 2  \ t         } e_3 + t e_4 \ , 
\label{Eq:SolDouble}
\eea
where $\alpha_i$ and $\beta_i$ are kinematical factors determined by the cut conditions. The coefficients
can be extracted from the large-$t$ expansion,
\bea
\label{Eq:Double1}
\left [  {N_+ \over \prod_{h \neq i,j} \db{h, +} } 
- \sum_{k \neq i,j }^{n-1}
         { \Delta^R_{ i j k , \; +} \over 
           \db{k, +} } \right ] \Bigg  |_{t\to \oo} 
&=& c_{2,0}^{(i j)} +c_{2,9}^{(i j)} \mu^2 +   c_{2,1}^{(i j)}  \ x_1
+c_{2,2}^{(i j)} \  x_1^2 + \nn
&-&   c_{2,5}^{(i j)}  \ t +  c_{2,6}^{(i j)} \ t^2 -  c_{2,8}^{(i j)} \ x_1 t  + \mathcal{O} \left ( {1 \over t }\right ) \  , \\
\left [ {N_- \over \prod_{h \neq i,j} \db{h, -} }
- \sum_{k \neq i,j }^{n-1}
         { \Delta^R_{ i j k , \; -} \over 
           \db{k, -} } \right ]\Bigg |_{t\to \oo} 
&=& c_{2,0}^{(i j)} +c_{2,9}^{(i j)} \mu^2 +   c_{2,1}^{(i j)}  \ x_1
+c_{2,2}^{(i j)} \  x_1^2 + \nn
&-&    c_{2,3}^{(i j)}  \ t  +  c_{2,4}^{(i j)}  \ t^2 -  c_{2,7}^{(i j)} \ x_1 t   + \mathcal{O} \left ( {1 \over t }\right )\ .
\label{Eq:Double2}
\eea
Eqs.~(\ref{Eq:Double1})  and~(\ref{Eq:Double2}) hold only if the uncut denominators are 
linear in $t$, namely
\bea
D_{h,\pm} \stackrel{t\to\infty}{=}  2 e_{3,4} \cdot (p_h - p_i) \ t  + {\cal O}( 1)   \quad  \forall \; h\ne i,j.
\eea
Therefore the momentum  $K_2$, entering in the definition of $e_{3,4}$, has to be chosen so that 
\bea
(p_h -p_i) \cdot e_{3,4} \neq 0 
\label{Eq:CondBT}
\eea
for all  $ h\ne i,j$. \\

The terms involving  the reduced residues $\Delta^R_{ijk}$ remove 
the contributions of the spurious three-points coefficients. The treatment 
of the subtraction terms is thus different from the one proposed 
in~\cite{Forde:2007mi,Kilgore:2007qr,Badger:2008cm}, where 
the spurious  three-points contamination is removed by subtracting all possible triple cuts 
constructed from the double cut $(ij)$. \\

We remark that in general neither
\bea
   {N_\pm \over \prod_{h \neq i,j} \db{h, \pm} } \qquad \mbox{nor} \qquad
         { \Delta^R_{ i j k , \; \pm} \over 
           \db{k, \pm} } \nonumber
\eea 
are polynomial in $t$ and $1/t$, but only their difference so it is. Instead 
their Laurent expansion has the same polynomial structure of  the r.h.s. 
of eqs.~(\ref{Eq:Double1}) and~(\ref{Eq:Double2}). For the ``$+$'' case we have
\bea
  {N_+ \over \prod_{h \neq i,j} \db{h, +} }  \Bigg  |_{t\to \oo} &=& a_{2,0}^{(i j)} +a_{2,9}^{(i j)} \mu^2 +   a_{2,1}^{(i j)}  \ x_1
+a_{2,2}^{(i j)} \  x_1^2 +\nn
&&-   a_{2,5}^{(i j)}  \ t +  a_{2,6}^{(i j)} \ t^2 -  a_{2,8}^{(i j)} \ x_1 t   + \mathcal{O} \left ( {1 \over t }\right ) \  ,  \\
%
{ \Delta^R_{ i j k , \; +} \over 
           \db{k, +} } \Bigg  |_{t\to \oo} 
&=& b_{2,0}^{(i j | k)} +b_{2,9}^{(i j | k)} \ \mu^2 +   b_{2,1}^{(i j | k)}  \ x_1
+b_{2,2}^{(i j | k)} \  x_1^2 + \nn
&& -   b_{2,5 }^{(i j  | k)}  \ t +  b_{2,6}^{(i j | k)} \ t^2 -  b_{2,8}^{(i j | k)} \ x_1 t   + \mathcal{O} \left ( {1 \over t }\right ) \  .
\label{Eq:DecAB}
\eea
The ``$-$'' case is obtained by replacing $(5,6,8) \to (3,4,7)$.
Therefore in our algorithm the coefficients $a^{(ij)}$ and $b^{(ij | k)}$ can be computed {\it separately}, 
obtaining the coefficient $c^{(ij)}$ by their difference,
\bea
c^{(ij)}_{2,m} = a^{(ij)}_{2,m} - \sum_{k \neq i,j }^{n-1} b^{(ij | k)}_{2,m} \ . 
\label{Eq:DoubleSUB} 
\eea
In other words the subtraction is implemented at the coefficient-level rather than at the integrand-level. 
Moreover, given the known structure of $\Delta^R_{ijk}$, 
the analytic expression of the coefficients $b^{(ij | k)}$ can be computed once and for all, in terms of 
the 3-point spurious coefficients, the corresponding basis, and the basis of the cut $(ij)$.
The actual semi-numerical procedure has to be applied {\it only} to the term involving the numerator,
in order to determine the coefficients $a^{(ij)}$.

\subsection*{Single cut}
We consider the  following solution of the single cut  $\db{i} = 0$,
\bea
q^{(i)}_+ &=& - p_i + x_1 e_1 + { \alpha_0 + \mu^2 \over 2  x_1 } e_2 
\label{Eq:SingleS}
\eea
with $\alpha_0$ fixed  by the cut conditions.  The coefficient $c^{(i)}_{1,0}$
is extracted from the large-$x_1$ limit,
\bea
\left [ {N_+ \over \prod_{h \neq i} \db{h, +} } 
- \sum_{j<k \neq i }^{n-1}
         { \Delta^R_{ i j k , \; +} \over 
           \db{j, +}   \db{k, +}  } 
-\sum_{j \neq i }^{n-1}
         { \Delta^R_{i j , \; +} \over 
           \db{j, +}  } \right ] \Bigg |_{x_1 \to \oo} =    c_{1,0}^{(i)} 
 + \mathcal{O}\left ({ 1 \over x_1 } \right ) + \Omega \left (x_1  \right ) \, . \ \
\label{Eq:Single1}
\eea
The symbol  $\Omega(x_1)$ denotes terms which are not negligible with respect to $x_1$ as $x_1 \to \oo$.
Eq.~(\ref{Eq:Single1})  holds only if the uncut denominators are linear in $x_1$,
\bea
D_{h,+} \stackrel{x_1\to\infty}{=}  2 e_{1} \cdot (p_h - p_i) \ x_1  + {\cal O}( 1) \ ,
\qquad \mbox{ for } \; h\ne i \ . 
\eea
Therefore $K_1$ and $K_2$, entering the definition of the basis $e_{1,2}$, have to be chosen 
accordingly.  \\

The contributions from the spurious two- and three-points
coefficients are discarded   subtracting the reduced residues $\Delta^R_{ij}$ and 
$\Delta^R_{ijk}$. The subtraction procedure differs from the one presented in~\cite{Kilgore:2007qr},
where the spurious 2- and 3-point contributions are removed 
subtracting the double and triple cuts constructed from the single cut $(i)$. \\

Also in this case we remark that
\bea
   {N_+ \over \prod_{h \neq i} \db{h, +} } \ , \qquad 
         { \Delta^R_{ i j k , \; +} \over 
           \db{j, +}   \db{k, +}  }  \ , \qquad
\mbox{and} \qquad
         { \Delta^R_{i j , \; +} \over 
           \db{j, +}  } \nonumber 
\eea 
are not separately polynomial in $x_1$ and $1/x_1$, but only their combination in Eq.(\ref{Eq:Single1}) so it is.
Instead, their Laurent expansion  has the same polynomial structure of the r.h.s. 
of eqs.~(\ref{Eq:Single1}),
\bea
   {N_+ \over \prod_{h \neq i} \db{h, +} }\Bigg|_{x_1 \to \oo}  &=&  a_{1,0}^{(i)}  + \mathcal{O}\left ({ 1 \over x_1 } \right ) + \Omega \left (x_1  \right ) \ , \\
%
         { \Delta^R_{ i j k , \; +} \over 
           \db{j, +}   \db{k, +}  }\Bigg|_{x_1 \to \oo}  &=&  b_{1,0}^{(i | jk)} + \mathcal{O}\left ({ 1 \over x_1 } \right ) + \Omega \left (x_1  \right ) \ , \\
         { \Delta^R_{i j , \; +} \over 
           \db{j, +}  }\Bigg|_{x_1 \to \oo}              &=&  b_{1,0}^{ (i | j)}  + \mathcal{O}\left ({ 1 \over x_1 } \right ) + \Omega \left (x_1  \right )  \ .  
\eea 
The coefficients $a^{(i)}$,  $b^{(i | jk)}$, and $b^{ (i | j )}$ can be computed  separately, and finally 
the 1-point coefficient read,
\bea
c^{(i)}_{1,0} = a^{(i)}_{1,0} - \sum_{j<k \neq i }^{n-1} b^{(i | jk)}_{1,0}  - \sum_{j \neq i }^{n-1} b^{ (i | j)}_{1,0} \ .
\eea
The semi-numerical procedure has to be used to  
determine the coefficient $a^{(i)}_{1,0}$  only.
Indeed, given the known structure of $\Delta^R_{ijk}$, and $\Delta^R_{ij}$, 
the parametric form of the coefficients $b^{(i | jk)}_{1,0}$ and $b^{(i | j)}_{1,0}$ is universal and 
can be computed once and for all.

%% file: implementation.tex
\def\var{\tau}
\def\coe{s}

\section{Implementation}
\label{Sec:PolyDiv}
As shown in the previous section, our method requires, from quadruple- to single-cut,  
one-dimensional asymptotic expansions. 
In this section we present their semi-numerical implementation.

\subsection*{Quadruple cut}
The coefficient $c^{(ijk\ell)}_{4,4}$ is computed performing a large $\mu^2$ expansion
of 
\bea
{N_{+}
\over \prod_{h \neq i,j,k,\ell}^{n-1} D_{h, \; +}
},
\eea
cfr. eq.~(\ref{Eq:Quad2}). Both $N$ and $D_h$ are polynomial in $u$
\bea
N_{+} = \sum_{i=1}^r f_i \ u^i = N\left ( u v_\perp, v^2_\perp u^2 \right ) + \mathcal{O}(u^{r-1})  , \qquad  
D_{h, \; +} = d_{h,0} + d_{h,1} u \ . 
\eea
By power counting, the coefficient $c^{(ijk\ell)}_{4,4}$ is non-vanishing only if $r=n$ and it is proportional
to $f_r$,
\bea
c^{(ijk\ell)}_{4,4} = {f_r \over (v^2_\perp)^r \prod_{h\neq i,j,k,\ell}^{n-1} d_{h,1} } = 
{f_r \over (v^2_\perp)^r \prod_{h\neq i,j,k,\ell}^{n-1} (2 p_h \cdot v_\perp) } \ .
\eea
The coefficient $f_r$ can be obtained from the analytic expression of $N  (u v_\perp, v_\perp^2 u^2)$.
For instance this procedure can be easily implemented in {\Formcalc} and {\Gosam}
which use the symbolic  manipulations  programs  {\mathematica} and/or {\form}~\cite{Vermaseren:2000nd}.

\subsection*{Triple, double, and single cuts}
Along the reduction procedure, the integrand of the $n$-ple cut is a multivariate function of $(5-n)$ variables,
corresponding to the parameters of the loop momentum not fixed by the on-shell conditions.
Each expansion is performed with respect to one variable only, say $\var$. 
The solution of the triple, of the double and of the single cut reads
as follows
\bea
  q_{\rm{cut}}^\mu = \frac{1}{\var}\, \eta_{-1}^\mu + \eta_0^\mu + \var \, \eta_1^\mu \ , 
\eea
in terms of the (cut-dependent) momenta $\eta_{-1}$, $\eta_0$, and $\eta_1$.
For each cut the generic term to be expanded is a ratio of the type,
\bea
{F(\var) \over \prod_{h=0}^{k-1} D_h(\var)},
\label{Eq:RatioG}
\eea
where $D_h$ is an  uncut propagators,
\bea
D_h(\var) = { \mathcal{D}_h(\var) \over \var},  \qquad   \mathcal{D}_h(\var) =  \sum_{i=0}^2 d_{h,i} \; \var^i .
\eea
The function $F$ can either be the original numerator $N$ or any of the reduced residues $\Delta^R$.

If $F$ is a reduced residue $\Delta^R$ the large $\var$ expansion of eq.~(\ref{Eq:RatioG}) is universal
and can be  performed analytically, cfr. section~\ref{SSec:MethodCoeff}.

If $F$ is the numerator $N$ we have
\bea
F(\var) = { \mathcal{F}(\var) \over \var^r },   \qquad  \mathcal{F}(\var)  = \var^r  N\left( q_{\rm{cut}} ,  \mu^2 \right )    \equiv \sum_{i=0}^{2r} f_i \; \var^i  , 
\label{Eq:DefF}
\eea
where $r$ is the rank of the numerator.  
The coefficients $f_i$  depend implicitly on the cut through the momenta 
$\eta_{-1}$, $\eta_0$, and $\eta_1$, 
\bea
f_i = f_i\left (\eta_{-1}, \eta_{0}, \eta_{1} \right ).
\eea
Their parametric  expression  can be obtained from either  the tensor structure of the integrand 
or the   analytic form  of the  numerators,  as provided by codes like  {\Formcalc} and {\Gosam}. 
In the $F=N$ case the large $\var$ expansion is numerically  implemented through polynomial division,
as described below.  In the following we assume $r \ge k$, otherwise 
the ratio (\ref{Eq:RatioG}) vanishes in the $\var \to \infty$ limit.

\paragraph{Step 1.} We  start by dividing $\mathcal{F}$ by $\mathcal{D}_0$ obtaining
\bea
{\mathcal{F}(\var) \over \mathcal{D}_0(\var)} =  \mathcal{Q}_0(\var)  + { \mathcal{R}_0(\var) \over  \mathcal{D}_0 } \, .
\eea
The quotient $\mathcal{Q}_0$ is a polynomial of degree $2r-2$, while 
the remainder $\mathcal{R}_0$ is a polynomial of degree one. In the large-$\var$ limit, the contribution of 
the latter can be neglected, since 
\bea
{\mathcal{R}_0(\var) \over \prod_{h=0}^{k-1} \mathcal{D}_h(\var)}  \stackrel{\var \to\infty}{=}  \mathcal{O}\left ( {1 \over \var^{2k-1}} \right ) \ , 
\eea
therefore
\bea
{\mathcal{F}(\var) \over \prod_{h=0}^{k-1} \mathcal{D}_h(\var)}  \stackrel{\var \to\infty}{=} {\mathcal{Q}_0(\var) \over  \prod_{h=1}^{k-1} \mathcal{D}_h(\var)} \ .
\eea
\paragraph{Step 2.} We perform the division by the successive denominator $\mathcal{D}_1$
\bea
{\mathcal{Q}_0(\var) \over \mathcal{D}_1(\var)} =  \mathcal{Q}_1(\var)  + { \mathcal{R}_1(\var) \over  \mathcal{D}_1 } \, .
\eea
The quotient  $\mathcal{Q}_1$ is a polynomial of degree $2r-4$ while the remainder $\mathcal{R}_1$ has degree one. In the large-$\var$ limit, the contribution of $\mathcal{R}_1$ drops out as well, hence 
\bea
{\mathcal{F}(\var) \over \prod_{h=0}^{k-1} \mathcal{D}_h(\var)}  \stackrel{\var \to\infty}{=} {\mathcal{Q}_1(\var) \over  \prod_{h=2}^{k-1} \mathcal{D}_h(\var)} \ .
\eea 

\paragraph{Last step. 
}

After reiterating this procedure over the remaining denominators, $\mathcal{D}_2, \ldots, \mathcal{D}_{k-1}$, 
we get 
\bea
{\mathcal{F}(\var) \over \prod_{h=0}^{k-1} \mathcal{D}_h(\var)}  \stackrel{\var \to\infty}{=} \mathcal{Q}_{k-1}(\var) \equiv \sum_{i=0}^{2(r-k)} \coe_{i} \; \var^i \ ,
\eea
Finally, the Laurent expansion of eq.~(\ref{Eq:RatioG}) is given by 
\bea
{F(\var) \over \prod_{h=0}^{k-1} D_h(\var)} =  
{\mathcal{F}(\var) \over \var^{r-k}  \prod_{h=0}^{k-1} \mathcal{D}_h(\var)}
 \stackrel{\var \to\infty}{=} 
\sum_{i=0}^{r-k} \coe_{i+r-k}\;  \var^i + \mathcal{O}\left ( {1 \over \var} \right ) \ .
\label{Eq:LaurentFinal}
\eea
It  is worth to notice that the large $\tau$ expansion can be achieved by using a smaller polynomial
\bea
\mathcal{F}^R(\var) = \sum_{i=r+k}^{2r} f_i \var^i \ ,  
\label{Eq:DefFR}
\eea
instead of the polynomial $\mathcal{F}$ defined in eq.~(\ref{Eq:DefF}). Indeed 
\bea
{F(\var) \over \prod_{h=0}^{k-1} D_h(\var)} \stackrel{\var \to\infty}{=} 
{\mathcal{F}^R(\var) \over \var^{r-k}  \prod_{h=0}^{k-1} \mathcal{D}_h(\var)} \stackrel{\var \to\infty}{=} 
\sum_{i=0}^{r-k} \coe_{i+r-k}\;  \var^i  + \mathcal{O}\left ( {1 \over \var} \right )  \ .
\label{Eq:LaurentFinal2}
\eea

We have implemented this algorithm in {\C++} and in {\mathematica} and verified its correctness 
reconstructing  the integrands of up to sixth rank 6-point functions. Two numerical examples are 
described in the Appendix~\ref{App:Nume}. A complete implementation in {\Gosam} and {\Formcalc} is planned.

%% file: example.tex
\renewcommand{\Re}{\mathrm{Re}}
\renewcommand{\Im}{\mathrm{Im}}
\def\Res{\mathrm{Res}}
\def\lp{\left(}
\def\rp{\right)}

\def\A{\mathcal{A}}
\def\S{\mathcal{S}}
\def\O{\mathcal{O}}
\def\nn{\nonumber \\}

\def\la{\langle}
\def\ra{\rangle}
\def\lb{[}
\def\rb{]}
\def\spa#1{\langle#1\rangle}
\def\spb#1{[#1]}
\def\spab#1{\langle#1]}
\def\spba#1{[#1\rangle}

\def\hi{\hat{i}}
\def\hj{\hat{j}}
\def\hQ{\hat{Q}}
\def\samurai{{{\sc samurai}}}

\def\vi{v}
\def\wu{w}

\section{Example: reducing a second rank  3-point integrand}
\label{sec:anexample}
In this section we apply the reduction procedure described above 
considering a rank-two three-point integrand of the type
\begin{equation}
  \label{eq:rank2tr}
  \frac{N(q)}{D_0 D_1 D_2}\equiv \frac{4 (q\cdot \vi)(q\cdot \wu)}{D_0 D_1 D_2} \ ,
\end{equation}
where
\begin{equation}
  \label{eq:trden}
  D_0 = q^2-m^2, \qquad 
  D_1 = (q-k_1)^2-m^2 , \qquad  
  D_2 = (q+k_2)^2-m^2 \ .  
\end{equation}
the ``external'' momenta $k_1$ and $k_2$ taken as massless.
For simplicity we consider only the four-dimensional part of the reduction.
The extension to $d$-dimensions is straightforward.  
For illustration purposes, we use polynomial division .

\subsection*{The  cut $\mathbf{(012)}$}
In order to deal with relatively compact expressions we use the basis  $\{e_1, e_2, e_3, e_4 \}$, where
\bea
e_1^\mu = k_1^\mu , \qquad e_2^\mu = k_2^\mu, \qquad   e_3^\mu =  \frac{\spab{1| \gamma^\mu| 2}}{2}, \qquad 
 e_4^\mu =  \frac{\spab{2| \gamma^\mu| 1}}{2}.
\label{Eq:Dummy}
\eea
The basis  does not fulfill the normalization 
conditions $e_1 \cdot e_2 = - e_3 \cdot e_4 =1$. Therefore  
the formulae~(\ref{Eq:Triple1}--\ref{Eq:Triple3})  have to be modified performing the substitutions
\begin{align*}
c_{3,i} & \to (e_1\cdot e_2) c_{3,i},    &\mbox{ if } & i = 1,4,8,9 \ ;  \nn
c_{3,i} & \to (e_1\cdot e_2)^2 c_{3,i},  &\mbox{ if } &  i = 2,5  \ ; \\
c_{3,i} & \to (e_1\cdot e_2)^3 c_{3,i},  &\mbox{ if } &  i = 3,6  \ . 
\end{align*}
The solutions of the triple cut are
\bea
   q^{(012)}_+ = t \ e_3 -
\frac{m^2}{2(k_1\cdot k_2) \ t} \ e_4 \ , \quad
   q^{(012)}_- = 
t \ e_4 -
\frac{m^2}{2(k_1\cdot k_2) \ t} e_3 \ .
\label{eq:trhpar}
\eea
The functions appearing the l.h.s.\ of eqs.~(\ref{Eq:Triple1}--\ref{Eq:Triple3}) are
\begin{align}
  N_{+} = -\frac{m^2}{2(k_1\cdot k_2)}\Big(
\spab{1|\vi|2}\,\spab{2|\wu|1}+\spab{2|\vi|1}\,\spab{1|\wu|2} \Big) + \spab{1|\vi|2}\,\spab{1|\wu|2}\, t^2 + \O\left ( {1 \over t^2} \right ) , \\
  N_{-} = -\frac{m^2}{2(k_1\cdot k_2)}\Big(
\spab{1|\vi|2}\,\spab{2|\wu|1}+\spab{2|\vi|1}\,\spab{1|\wu|2} \Big) + \spab{2|\vi|1}\,\spab{2|\wu|1}\, t^2 + \O\left ({1\over t^2} \right ),
\end{align}
since all the propagators are cut. No polynomial division is needed and the coefficients 
can be immediately computed. The non vanishing ones are  
  \begin{align}
    & c_{3,0}^{(012)}  = -\frac{m^2}{2(k_1\cdot k_2)}\Big(\spab{1|\vi|2}\,\spab{2|\wu|1}+\spab{2|\vi|1}\,\spab{1|\wu|2} \Big) , \\
    & c_{3,2}^{(012)}   = \frac{\spab{2|\vi|1}\,\spab{2|\wu|1}}{(k_1\cdot k_2)^2} , \\
    & c_{3,5}^{(012)}   = \frac{\spab{1|\vi|2}\,\spab{1|\wu|2}}{(k_1\cdot k_2)^2} . 
  \end{align}
The reduced residue reads as follows
\begin{equation}
\label{eq:tr2Ns3}
  \Delta^R_{012}(q) = c_{3,5}^{(012)} \lp \frac{\spab{2| q| 1}}{2} \rp ^2 + c_{3,2}^{(012)} \lp \frac{\spab{1| q| 2}}{2}\rp ^2 \ .
\end{equation}

\subsection*{The cut $\mathbf{ (21)}$}
The basis used for this cut is obtained from the momenta 
\bea
K^\mu_1= k^\mu_1+k^\mu_2 , \qquad K^\mu_2 = -\frac{\spab{1|\gamma^\mu|2}}{2}-\frac{\spab{2|\gamma^\mu|1}}{2} \ ,
\eea
and  its elements  are  
\bea
 e_1^\mu = \frac{1}{2}\left (  K_1^\mu - K_2^\mu \right ) , \quad
 e_2^\mu = \frac{1}{2} \left (K_1^\mu + K_2^\mu  \right ), \quad
e_3^\mu  = \frac{\spab{e_1|\gamma^\mu|e_2}}{2}, \quad
e_4^\mu  = \frac{\spab{e_2|\gamma^\mu|e_1}}{2} . \quad
\label{Eq:Basis12}
\eea
The element of the basis are not canonically normalized thus eqs.~(\ref{Eq:Double1}) and~(\ref{Eq:Double2}) have to be
modified performing the substitutions
\begin{align*}
c_{2,i} &\to (e_1\cdot e_2) c_{2,i},    &\mbox{ if } &  i = 1,3,5 \ ;  \nn
c_{2,i} &\to (e_1\cdot e_2)^2 c_{2,i},  &\mbox{ if } &  i = 2,4,6,7,8   \ .
\end{align*}
The solutions of the double cut $(12)$ are 
\bea
  \label{eq:trr2b12}
    q^{(21)}_+ =  -k_2 +  x_1\, e_1+(1-x_1)e_2  +t\, e_3 + \left ( -  {m^2 \over (2 k_1\cdot k_2)} + x_1 -x_1^2  \right ) \, \frac{1}{t} e_4 \ ,  \label{eq:trr2b12a} \\
    q^{(21)}_- =  -k_2 +  x_1\, e_1+(1-x_1)e_2  +t\, e_4 + \left ( -  {m^2 \over (2 k_1\cdot k_2)} + x_1 -x_1^2  \right ) \, \frac{1}{t} e_3 \ .  \label{eq:trr2b12b}  
\eea
The coefficients  $c_{2,0}^{(21)}$, $c_{2,1}^{(21)}$ and $c_{2,5}^{(21)}$ are obtained from eq.~(\ref{Eq:Double1}). The large $t$ expansion is 
obtained performing the polynomial division with respect to $t$, along the lines of section~\ref{Sec:PolyDiv}. 
\begin{itemize}

\item {\it Contribution of the reduced residue} -- The only reduced residue  entering the subtractions is 
\bea
  \frac{\Delta^R_{012,\; +}}{D_{0,\;+}} = b^{(21 | 0)}_{2,0} + b_{2,1}^{(21 | 0)} x_1 - b_{2,5}^{(21 | 0)} t \ .
\eea
The coefficients $ b^{(21 | 0)}$ are universal functions of the spurious coefficients and of kinematic invariants. 
In the case of a rank-2 three-point integrand  with  $p_0=0$  they read as follows
\bea
 b_{2,0}^{(21 | 0)} &=&  \frac{-1}{ 4   (e_3^{(21)} \cdot p_2   )^2} \Bigg \{ \Bigg [ 
\left [\left (p^2_2 +m_2^2-m_0^2 \right )  - 2 \alpha_0 \left  ( e_2^{(21)} \cdot  p_2 \right ) 
\right ] \left ( e_3^{(21)} \cdot  e_3^{(012)} \right )   c_{3,2}^{(012)} \pagebreak[1] \nn 
&& + 2 \left [   c_{3,1}^{(012)}  + 2 \, \alpha_0 
\left ( e_2^{(21)} \cdot  e_3^{(012)} \right ) c_{3,2}^{(012)} \right ] \left  ( e_3^{(21)} \cdot  p_2 \right )  \Bigg ]  \left  ( e_3^{(21)} \cdot  e_3^{(012)} \right ) \pagebreak[1]  \nn
&&   + \Big [c_{3,1} \to c_{3,4};  \; c_{3, 2} \to   c_{3,5} ; \;\;  e_3^{(012)} \to   e_4^{(012)} \Big ] \Bigg \} \pagebreak[1] \nn
&=&
 \frac{ 3 \la 2 | \vi | 1] \la 2 | \wu | 1]-\la 1 | \vi | 2] \la 1 | \wu | 2] }{4 (k_1\cdot k_2)} \ , \pagebreak[1] \nn
 b_{2,1}^{(21 | 0)} &=&  \frac{1}{ 2   (e_3^{(21)} \cdot p_2   )^2} \Bigg \{ \Bigg [  
\left    ( e_1^{(21)} \cdot p_2   + \alpha_1\;  e_2^{(21)} \cdot p_2 \right ) \left (e_3^{(21)} \cdot e_3^{(012)}   \right ) \pagebreak[1]  \nn
&& -2 \left  ( e_1^{(21)} \cdot e_3^{(012)}  + \alpha_1\;  e_2^{(21)} \cdot e_3^{(012)} \right   ) \left   (e_3^{(21)} \cdot p_2 \right  ) \Bigg ]   \; \left (e_3^{(21)} \cdot e_3^{(012)} \right  )
  c_{3,2}^{(012)} \pagebreak[1] \nn
&&  + \Big [ c_{3, 2} \to   c_{3,5} ; \;\;  e_3^{(012)} \to   e_4^{(012)} \Big ] \Bigg \} 
=\frac{\la 1 |\vi | 2] \la 1 | \wu | 2] - \la 2 | \vi | 1] \la 2 | \wu |
1]}{k_1\cdot k_2} \pagebreak[1] \nn
%
%
%
 b^{(21 | 0)}_{2,5} &=& {  c^{(012)}_{3,2} \left (e_3^{(21)} \cdot e_3^{(012)}  \right )^2 +  c^{(012)}_{3,5} \left (e_3^{(21)} \cdot e_4^{(012)}  \right )^2 \over 2   (e_3^{(21)} \cdot p_2   )  } \pagebreak[1]  \nn
&=& \frac{\la 1 | \vi | 2] \la 1 | \wu | 2] +\la 2 | \vi | 1] \la 2 | \wu | 1]}{4 (k_1 \cdot k_2 )} \ .
\eea

\item  {\it Contribution of the numerator} -- We perform the polynomial division to compute the
large $t$ expansion of the ratio
\bea
  \frac{N_+ }{D_{0,\;+}} =    \frac{\mathcal{F}(t) }{t \ \mathcal{D}_0(t)}  \stackrel{t\to \oo}{=}  \frac{\mathcal{F}^R(t) }{t \ \mathcal{D}_0(t)}  \ .
\eea
The function $\mathcal{F}$ defined in eq.~(\ref{Eq:DefF}) is given by $\mathcal{F} = t^2 N_+$.   
The reduced polynomial $\mathcal{F}^R$, eq.~(\ref{Eq:DefFR}), is obtained  from   $\mathcal{F}$
neglecting terms of $\O(t^2)$,
\bea
   \mathcal{F}^R(t) = f_4 t^4 + f_3 t^3 \ ,
\eea
where
\begin{align*}
  f_4 & = \frac{1}{4} \bigg(\la 1|\vi|1]-\la 1|\vi|2]+\la 2|\vi|1]-\la
2|\vi|2]\bigg)\bigg(\la 1|\wu|1]-\la 1|\wu|2]+\la 2|\wu|1]-\la
2|\wu|2]\bigg) \pagebreak[1] \nn
  f_3 & = \frac{1}{2} \bigg[ \Big (\la 1|\vi|1]-\la 1|\vi|2]-\la 2|\vi|2] \Big )  \Big (\la 1|\wu|1]-\la 1|\wu|2]-\la 2|\wu|2] \Big ) 
-\la 2|\vi|1] \la 2|\wu|1]\bigg ] \pagebreak[1] \nn
  & \quad +\frac{1}{2} \bigg [ \Big (\la 1|\vi|1]-\la 1|\vi|2]+\la 2|\vi|1]-\la
2|\vi|2] \Big ) \Big (\la 1|\wu|2]+\la 2|\wu|1] \Big )+\Big (\vi\leftrightarrow \wu \Big)\bigg ] x_1. \pagebreak[1] \nonumber
\end{align*}
The polynomial $\mathcal{D}_0$ in the denominator is related to the propagator
$D_{0, \; +}$,
\bea
\mathcal{D}_0(t) = t \  D_{0 \; , +} = d_{0,2}t^2 + d_{0,1}t + d_{0,0} \ ,
\eea
with 
\bea
 d_{0,2} = d_{0,1} = - (k_1 \cdot k_2) \ , \quad d_{0,0} = 
(k_1 \cdot k_2)  \lp \frac{m^2}{2 k_1\cdot k_2} - x_1 + x_1^2\rp \ .
\eea
In the notation of eq.~(\ref{Eq:LaurentFinal}), the result of the polynomial division reads, 
\bea
\frac{\mathcal{F}^R(t) }{ \mathcal{D}_0(t)} = \coe_2 t^2 + \coe_1 t + \coe_0 
+ { {\cal R}_0 \over{\cal D}_0  }
\quad  \Longrightarrow \quad  \frac{N_+ }{D_{0,\;+}}  \stackrel{t\to \oo}{=}\frac{\mathcal{F}^R(t) }{ t \mathcal{D}_0(t)}  \stackrel{t\to \oo}{=}  \coe_2 t + \coe_1 \ ,
\label{Eq:ExDiv}
\eea
where
\bea
  \coe_2 = {f_4 \over d_{0,2}} \ , \qquad \mbox{and} \qquad
  \coe_{1} =  {f_3 d_{0,2} - f_4 d_{0,1} \over d_{0,2}^2} \equiv \coe_{1,0} + x_1 \coe_{1,1} \ .
\eea
By comparing eqs.~(\ref{Eq:DecAB}) and~(\ref{Eq:ExDiv}) we get 
\bea
a_{2,0}^{(21)} = \coe_{1,0} \ , \qquad a_{2,1}^{(21)} = {  \coe_{1,1} \over (k_1\cdot k_2)} \ , \qquad a_{2,5}^{(21)} =  { -  \coe_2 \over (k_1\cdot k_2)}
\eea 
while $a_{2,2}^{(21)}$, $a_{2,6}^{(21)}$ and $a_{2,8}^{(21)}$ vanish. 
\item  {\it Computation of the coefficients} -- The coefficients $c^{(21)}$ are obtained subtracting the coefficients $b^{(21 | 0)}$ 
to the coefficients  $a^{(21)}$, according to eq.~(\ref{Eq:DoubleSUB}). The non-vanishing ones are:
\bea
 c_{2,0}^{(21)}   &=& \frac{1}{4 (k_1\cdot k_2)} \Bigg(\la 1|\vi|2]\, \la 1|\wu|1]     +\la 1|\wu|1]\, \la 2|\vi|1]-\la 1|\wu|2]\, \la 2|\vi|1] \pagebreak[1] \nn
&&  +\la 1|\wu|1]\, \la 2|\vi|2]     -\la 1|\wu|2]\, \la 2|\vi|2]     -\la 2|\vi|2]\, \la 2|\wu|1] \pagebreak[1] \nn
&&    - { \la 2|\vi|2]\, \la 2|\wu|2] \over 2}  - { \la 1|\vi|1]\, \la 1|\wu|1] \over 2 }\Bigg)  + \Big ( \vi \leftrightarrow \wu  \Big )  \ , \pagebreak[1] \nn
 c_{2,1}^{(21)} &=&   \frac{1}{2 (k_1\cdot k_2)^2} \; \Big(\la 1|\wu|2]\, \la 2|\vi|2] - \la 1|\vi|2]\, \la 1|\wu|1]- \la 1|\wu|1]\la 2|\vi|1] \pagebreak[1] \nn
&&      +  \la 2|\vi|2]\, \la 2|\wu|1] \Big) + \Big ( \vi \leftrightarrow \wu  \Big )  \ .  \pagebreak[1]  \nn
c_{2,5}^{(21)}  &=& \frac{-1}{4(k_1\cdot k_2)^2}\Bigg( \la 1|\vi|2]\, \la
1|\wu|1]-\la 1|\wu|1]\, \la 2|\vi|1]+\la 1|\wu|2]\, \la 2|\vi|1] +\la
1|\wu|1]\, \la 2|\vi|2] \pagebreak[1] \nn
&& -\la 1|\wu|2]\, \la 2|\vi|2] +\la 2|\vi|2]\, \la 2|\wu|1] -\frac{\la
1|\vi|1]| \la 1|\wu|1]}{2}-\frac{\la 2|\vi|2]\, \la 2|\wu|2]}{2}\Bigg) \pagebreak[1] \nn
&& + \Big( \vi \leftrightarrow \wu \Big) \ . \pagebreak[1] \nonumber
\eea

\end{itemize}

\noindent
The remaining non-vanishing coefficient, $c_{2,3}^{(21)}$, is 
obtained in a similar way, using eq.~(\ref{Eq:Double2}). The outcome is 
\bea
c_{2,3}^{(21)} &=& -\frac{1}{4(k_1\cdot k_2)^2}\Bigg(  \la 1|\wu|1]\, \la 2|\vi|1]+\la 1|\wu|2]\, \la 2|\vi|1]+\la 1|\wu|1]\, \la 2|\vi|2] \nn
&& +\la 1|\wu|2]\, \la 2|\vi|2] -\la 2|\vi|2]\, \la 2|\wu|1] -\la 1|\vi|2]|\la 1|\wu|1]  \nn
&&  - { \la 2|\vi|2]\, \la 2|\wu|2] \over 2}   - { \la 1|\vi|1]|\la 1|\wu|1] \over 2 } \Bigg)  + \Big ( \vi \leftrightarrow \wu  \Big )    \ .
\eea
The reduced polynomial $\Delta_{12}^R$ is given by
\bea
  \Delta_{21}^R &=& 
{ c_{2,1}^{(21)} +c_{2,3}^{(21)} +c_{2,5}^{(21)} \over 2} (q+k_2)\cdot k_1 +
{ c_{2,1}^{(21)} -c_{2,3}^{(21)} -c_{2,5}^{(21)} \over 2} (q \cdot k_2) \nn
&+& { c_{2,5}^{(21)} -c_{2,3}^{(21)} -c_{2,1}^{(21)} \over 2} { \la 1|q|2] \over 2}+
{ c_{2,3}^{(21)} -c_{2,5}^{(21)} -c_{2,1}^{(21)} \over 2} {\la 2|q|1] \over 2} 
\eea

\subsection*{The cut $\mathbf{(02)}$}
The computation of the coefficients of the cut $(02)$
proceeds along the lines described above. We define the basis 
using  
\bea
K^\mu_1 = k^\mu_2, \qquad K_2^\mu= k_1^\mu +\frac{\spab{1|\gamma^\mu|2}}{2}+\frac{\spab{2|\gamma^\mu|1}}{2} \ ,
\eea
and
\bea
 e_1^\mu =   K_1^\mu , \quad
 e_2^\mu = \left (K_1^\mu + K_2^\mu  \right ), \quad
e_3^\mu  = \frac{\spab{e_1|\gamma^\mu|e_2}}{2}, \quad
e_4^\mu  = \frac{\spab{e_2|\gamma^\mu|e_1}}{2} . \quad
\eea
The solutions of the cut are:
\bea
q^{(02)}_+ = x_1\, e_1 + t\, e_3 -\frac{m^2}{2(k_1\cdot k_2)\, t} e_4  \ , \qquad 
q^{(02)}_- = x_1\, e_1 -\frac{m^2}{2(k_1\cdot k_2)\, t} e_3 + t\, e_4 \ .
\eea
The non-vanishing coefficients are 
  \begin{align}
    & c_{2,1}^{(0 2)} = -\frac{\la 2|\vi|2]\, \la 2|\wu|2]}{2 (k_1\cdot k_2)^2} \\
    & c_{2,5}^{(0 2)} = \frac{\la 2|\vi|2]\, \la 2|\wu|1]+\la 2|\vi|1]\, \la 2|\wu|2]+\la 2|\vi|2]\, \la 2|\wu|2]}{2 (k_1\cdot k_2)^2} \\
    & c_{2,3}^{(0 2)} = \frac{\la 1|\wu|2]\, \la 2|\vi|2]+\la 1|\vi|2]\, \la 2|\wu|2]+\la 2|\vi|2]\, \la 2|\wu|2]}{2 (k_1\cdot k_2)^2}, 
  \end{align}
while  the reduced residue reads as follows
\bea
  \Delta_{02}^R &=& c_{2,1}^{(0 2)} q \cdot k_1  + \left(  c_{2,1}^{(0 2)} + c_{2,3}^{(0 2)} + c_{2,5}^{(0 2)} \right )  q \cdot k_2 \nn 
&+& { c_{2,1}^{(0 2)} +  c_{2,5}^{(0 2)}  \over 2 }   \la 1|q|2] +  { c_{2,1}^{(0 2)} +  c_{2,3}^{(0 2)}  \over 2 }   \la 2|q|1]
\eea

\subsection*{The cut $\mathbf{(2)}$}
We parametrize the single cut solution~(\ref{Eq:SingleS}) in the basis~(\ref{Eq:Basis12}),
\begin{equation}
  q^{(2)}_+ = - k_2 + x_1\, e_1 + \frac{m^2}{(2 k_1\cdot k_2)}\frac{1}{x_1} e_2.
\end{equation}
The large $x_1$ expansion in eq.~(\ref{Eq:Single1}) is obtained from the large $x_1$ 
expansion of the subtraction coefficients 
\bea
\frac{\Delta^R_{012, \; +}}{D_{0,\; +}D_{1,\; +}}  \Bigg  |_{x_1\to \oo} &=&   b^{(2 | 01)}_{1,0} +\mathcal{O}\left ({1 \over x_1} \right ) + \Omega (x_1)  \, , \nn
\frac{\Delta^R_{21,\; +}}{D_{1,\; +}}           \Bigg  |_{x_1\to \oo} &=&   b^{(2 | 1)}_{1,0} +\mathcal{O}\left ({1 \over x_1} \right ) +  \Omega (x_1) \, , \nn
\frac{\Delta^R_{02,\; +}}{D_{0,\; +}}           \Bigg  |_{x_1\to \oo} &=&   b^{(2 | 0)}_{1,0} +\mathcal{O}\left ({1 \over x_1} \right ) +  \Omega (x_1) \, , 
\eea
and from the polynomial division  of the ratio
\bea
  \frac{N_+}{D_{0,\; +}D_{1,\; +}}   =  a^{(2)}_{1,0} +\mathcal{O}\left ({1 \over x_1} \right ) \; .
\eea
The tadpole coefficients is
\bea
  c^{(2)}_{1,0} &=&  a^{(2)}_{1,0} -  b^{(2 | 01)}_{1,0} -  b^{(2 | 1)}_{1,0} -  b^{(2 | 0)}_{1,0}  = \frac{1}{8 (k_1\cdot k_2)^2}\Bigg(\la 1|\wu|2]\, \la 2|\vi|1] -\la 1|\wu|1]\, \la 2|\vi|1] \nn
&& + \la 1|\wu|1]\, \la 2|\vi|2]   +3 \la 1|\wu|2]\, \la 2|\vi|2] +3 \la 2|\vi|2]\, \la 2|\wu|1]  \nn
&& -\la 1|\vi|2] \la 1|\wu|1] + { \la 2|\vi|2]\, \la 2|\wu|2] \over 2 } + { \la 1|\vi|1]\la 1|\wu|1] \over 2 } \Bigg ) + \Big (\vi \leftrightarrow \wu  \Big )\ .
\eea

%% file: extension.tex
\section{Extended decomposition}
In the previous sections we assumed to deal with a renormalizable
theory, where the rank $r$ of the numerator can not be greater than
the number of external legs $n$.  In this section we extend the
integrand decomposition to the case where the rank $r$ can become
larger than $n$.

In general the residue of an $m$-point function is a 
a multivariate polynomial in $\mu^2$ and the ISP's 
characterizing  the residue. Each monomial has to be 
irreducible and its maximum rank has to be at most  $(m+r-n)$. 
In the following we list the irreducible monomial 
entering each cut. For later convenience 
we give  the decomposition of $g^{\mu \nu}$ in terms 
of the basis~(\ref{Eq:Eps}) and of the vectors~(\ref{Eq:VVp}),
\bea
\label{Eq:GMN1}
g^{\mu \nu} &=& \left (e_1^\mu e_2^\nu + e_2^\mu e_1^\nu  \right ) -  \left (e_3^\mu e_4^\nu + e_4^\mu e_3^\nu  \right ) \ ,  \\
g^{\mu \nu} &=& \left (e_1^\mu e_2^\nu + e_2^\mu e_1^\nu  \right ) + \frac{v^\mu \ v^\nu}{v^2} + \frac{v_\perp^\mu \ v_\perp^\nu}{v_\perp^2} \ . 
\label{Eq:GMN2}
\eea

\begin{itemize}
\item {\it Quintuple cut}, $(ijk\ell m)$ -- The only irreducible monomial is $\mu^2$. 
Indeed the residue of the quintuple cut does not have ISP's, thus
the allowed monomials are $(\mu^2)^\alpha$. Moreover from eq.~(\ref{Eq:GMN1})
\bea
\left ( \mu^2 \right )^\alpha  &=& \left [ D_i + m_i^2 - p_i^2  - 2\, (q\cdot p_i) - q^2 \right ]^\alpha \nn
 &=&   
\left [  D_i + m_i^2 - p_i^2  - 2\, (q\cdot p_i) - 2 \,  (q\cdot e_1) (q\cdot e_2) + 2\,  (q\cdot e_3) (q\cdot e_4) \right ]^\alpha \nn
&=& \mbox{constant terms} + \mbox{RSP's} \ ,
\eea
where the abbreviation ``RSP's'' means ``reducible scalar products''. 
This relation allows to express all the powers of  $\mu^2$ in terms of
a particular one, $(\mu^2)^{\alpha_0}$. As in the renormalizable case we choose
$\alpha_0 =1$ in order to decouple the contribution of the pentagons
from the computation of the coefficients of the boxes.

\item  {\it Quadruple cut}, $(ijk\ell)$ -- The irreducible monomials
are
\bea
  (\mu^2)^\alpha  \; \left (  (q+p_i)\cdot v_\perp \right )^\beta \qquad \mbox{ with } \beta = 0,1 \mbox{ and } \alpha=0,1,2, \ldots \ .
\eea
Eq.~(\ref{Eq:GMN2}) implies  
\bea
\left (  (q+p_i)\cdot v_\perp \right )^2 &=& v^2_\perp \, \left ( q^2 -  2 \, ((q + p_i )\cdot e_1)  ((q+p_i)\cdot e_2) -  \frac{ ((q+p_i)\cdot v)^2}{v^2}  \right ) \nn
&=& 
\mbox{constant terms} + \mbox{terms in }\mu^2 +\mbox{RSP's} \ , \quad \nonumber
%
\eea 
therefore the terms with  $\beta \ge 2$ are reducible.

\item   {\it Triple cut}, $(ijk)$ -- In this case the   irreducible monomials are 
\bea
  (\mu^2)^\alpha  \; \left (  (q+p_i)\cdot e_{3,4} \right )^\beta \qquad \mbox{ with } \alpha, \, \beta = 0,1,2, \ldots \ .
\eea
The monomials containing both $e_3$ and $e_4$ are reducible. Indeed from eq.~(\ref{Eq:GMN1}) 
\bea
\left (  (q+p_i)\cdot e_{3} \right ) \left (  (q+p_i)\cdot e_{4} \right ) = 
\mbox{constant terms} + \mbox{terms in }\mu^2 + 
\mbox{RSP's} \ . \quad \nonumber
\eea

\item    {\it Double cut}, $(ij)$ -- The  irreducible monomials
are of the type
\bea
  (\mu^2)^\alpha  \; \left (  (q+p_i)\cdot e_{3,4} \right )^\beta \;   \left (  (q+p_i)\cdot e_{2} \right )^\gamma  \qquad \mbox{ with } \alpha,  \beta,  \gamma = 0,1,2, \ldots \ .
\eea
As in the previous case, the monomials depending on both  $e_3$ and  $e_4$ are reducible. 

\item    {\it Single cut}, $(i)$ -- The  irreducible monomials
read as follows
\bea
  (\mu^2)^\alpha  \; \left (  (q+p_i)\cdot e_{1,2} \right )^\beta \;   \left (  (q+p_i)\cdot e_{3} \right )^\gamma   \left (  (q+p_i)\cdot e_{4} \right )^\delta  
\; \mbox{with } \alpha, \beta,  \gamma, \delta = 0,1, \ldots \; \; 
\eea
Eq.~(\ref{Eq:GMN1}) allows to write 
\bea
\left (  (q+p_i)\cdot e_{1} \right ) \left (  (q+p_i)\cdot e_{2} \right ) &=& \left (  (q+p_i)\cdot e_{3} \right ) \left (  (q+p_i)\cdot e_{4} \right ) \nn
&& +  \mbox{constant terms} + \mbox{terms in }\mu^2 + \mbox{RSP's} \ . \nonumber
\eea
Therefore  the terms containing both $e_1$ and $e_2$ do not enter the parametrization of the residue.
\end{itemize}

The residues $\Delta$ presented in section~\ref{Sec:IntegrandR} are the most general
polynomials with $r \le n$ satisfying these requirements.  Here we show,
as an example, their extension to the case $r \le n+1$.
In this case, the decomposition of the numerator has to be extended as follows:
\bea
\label{eq:2}
N(q,\mu^2) &=&
\sum_{i < \!< m}^{n-1}
          \Lambda_{ i j k \ell m}(q,\mu^2)
\prod_{h \ne i, j, k, \ell, m}^{n-1} \db{h} 
+\sum_{i < \!< \ell}^{n-1}
          \Lambda_{ i j k \ell }(q,\mu^2)
\prod_{h \ne i, j, k, \ell}^{n-1} \db{h} 
+ \pagebreak[1]  \nn     &+&
\sum_{i < \!< k}^{n-1}
          \Lambda_{i j k}(q,\mu^2)
\prod_{h \ne i, j, k}^{n-1} \db{h} 
+\sum_{i < j }^{n-1}
          \Lambda_{i j}(q,\mu^2) 
\prod_{h \ne i, j}^{n-1} \db{h} \nn
&+&\sum_{i}^{n-1}
          \Lambda_{i}(q,\mu^2) 
\prod_{h \ne i}^{n-1} \db{h} \pagebreak[1]  
+     \Lambda(q,\mu^2) 
\prod_{h=0}^{n-1} \db{h} 
 \ , \qquad
\label{def:OPP:decoext}
\eea
where the polynomials $\Lambda$ are defined as,
\bea
\Lambda_{ i j k \ell m}(q,\mu^2) &=&  \Delta_{ i j k \ell m}(q,\mu^2) \ , \pagebreak[1]  \nn [1.0ex]
\Lambda_{ i j k \ell}(q,\mu^2) &=&  \Delta_{ i j k \ell}(q,\mu^2) + c_{4,5}^{(ijk\ell)} \ \mu^4 \  (q+p_i)  \cdot v_\perp \ , \pagebreak[1]  \nn [1.0ex]
\Lambda_{ i j k }(q,\mu^2) &=&  \Delta_{ i j k}(q,\mu^2) + c_{3,14}^{(ijk)} \ \mu^4    + c_{3,10}^{(ijk)} \ \mu^2 \ \left ( (q+p_i)\cdot e_3 \right )^2 \nn
&&  + c_{3,11}^{(ijk)} \ \mu^2 \ \left ( (q+p_i)\cdot e_4 \right )^2  + c_{3,12}^{(ijk)} ( (q+p_i)\cdot e_3 )^4  \nn
&&  + c_{3,13}^{(ijk)} ( (q+p_i)\cdot e_4 )^4 \ ,  \pagebreak[1]  \nn [1.0ex]
\Lambda_{ i j }(q,\mu^2) &=&  \Delta_{ i j }(q,\mu^2) +  \mu^2 \Big (c_{2,10}^{(ij)} \  (q+p_i)\cdot e_2  + c_{2,11}^{(ij)} \ (q+p_i)\cdot e_3 \pagebreak[1]  \nn
&& + c_{2,12}^{(ij)} (q+p_i)\cdot e_4 \Big ) + c_{2,13}^{(ij)} \ ( (q+p_i) \cdot e_2)^3 + c_{2,14}^{(ij)} ( (q+p_i) \cdot e_3 )^3  \pagebreak[1]  \nn
&& + c_{2,15}^{(ij)} ( (q+p_i) \cdot e_4)^3 + c_{2,16}^{(ij)} ( (q+p_i) \cdot e_2)^2 ((q+p_i)\cdot e_3)  \pagebreak[1]  \nn
&& + c_{2,17}^{(ij)}((q+p_i)\cdot e_2)^2 ((q+p_i)\cdot e_4) \pagebreak[1]  \nn
&& + c_{2,18}^{(ij)}((q+p_i) \cdot e_2) ((q+p_i) \cdot e_3)^2 \pagebreak[1]  \nn
&& + c_{2,19}^{(ij)}((q+p_i)\cdot  e_2) ((q+p_i) \cdot e_4)^2 \ , \pagebreak[1]  \nn [1.0ex]
\Lambda_{ i  }(q,\mu^2) &=&  \Delta_{ i  }(q,\mu^2) + c_{1,5}^{(i)} ((q+p_i)\cdot e_1)^2 + c^{(i)}_{1,6}((q+p_i)\cdot e_2)^2 \pagebreak[1]  \nn
&&  + c^{(i)}_{1,7} ((q+p_i)\cdot e_3)^2+c^{(i)}_{1,8} ((q+p_i)\cdot e_4)^2 \pagebreak[1]  \nn
&&  + c_{1,10}^{(i)} ((q+p_i) \cdot e_1) ((q+p_i)\cdot e_3)    + c_{1,11}^{(i)}((q+p_i) \cdot e_1)((q+p_i) \cdot e_4) \pagebreak[1]  \nn
&&  + c_{1,12}^{(i)} (( q+p_i ) \cdot e_2)((q+p_i) \cdot e_3)  + c_{1,13}^{(i)} ((q+p_i) \cdot e_2 ) (( q+p_i )\cdot e_4) \pagebreak[1]  \nn
&&  + c_{1,14}^{(i)} \ \mu^2  + c_{1,15}^{(i)} ((q+p_i) \cdot e_3)((q+p_i) \cdot e_4) \ , \pagebreak[1]  \nn
\Lambda(q,\mu^2) &=&  c_{0} \ .
\label{def:Lambda}
\eea
The functions $\Delta$, appearing already in the case $r \le n$, were
given in Eqs.~(\ref{def:Resi5:right})--(\ref{def:Resi1:right}).
We observe that the polynomial residues of 4-, 3-, 2-, and 1-point function acquire a richer structure,
and a 0-point coefficient $c_0$ does appear.
The latter coefficient is needed for the complete reconstruction of the integrand, 
but it is spurious. Indeed it multiplies a scaleless integral, which  vanish in dimensional regularization.
The coefficient $c_0$ is not cut-constructible but it  can be computed by  inverting 
eq.~(\ref{def:OPP:decoext}) in correspondence to any 
value  of  $(q, \mu^2)$ not annihilating any  propagator.
 It can be shown that the 0-point coefficient is present in the decomposition of rank-two 
1-point integrals only. In the case of higher point integrals this term is absent, 
owing to mutual cancellation between different 1-point polynomials $\Lambda_i$ parametrised in 
terms of a common single-cut vector basis $\{ e_1,e_2,e_3,e_4 \}$.

We remark that according to our new algorithm, 
also in the case $r\le n+1$ the residues of the 5-point functions
and the spurious 4-point coefficients are not needed. 
Moreover, since the 0-point coefficient is spurious, 
 the spurious coefficients of the 1-point functions are not needed as well.
The other coefficients can be computed performing quadruple, triple, double and single cuts, using 
the series expansions described in section~\ref{SSec:MethodCoeff} and selecting the appropriate 
terms of the series.  The new coefficients  are obtained including  higher order contributions 
in the expansions. 
The nice features of the method hereby discussed are not spoiled by the presence of 
higher rank numerators, and the coefficients of the 4-point functions do not affect the determination
of the lower-point coefficients.
We checked the validity of our procedure reconstructing the integrands of up to seventh rank 6-point 
functions.

The one loop  $n$-point amplitude is obtained upon integration of Eq.(\ref{def:OPP:decoext}). 
The outcome reads,
\bea
{\cal A}_n +  \delta  {\cal A}_n \ , 
\eea
where ${\cal A}_n$ is given in eq.~(\ref{eq:Aresult}) and the new contribution  $\delta {\cal A}_n$ is
\bea
\delta {\cal A}_n &=& 
\sum_{i < j < k }^{n-1} 
  c_{3,14}^{ (i j k)} \ I_{i j k}[\mu^4] \pagebreak[1]  \nn
&+&
\sum_{i < j }^{n-1}\bigg\{
  c_{2,13}^{ (i j)} \ I_{i j}[((q+p_i)\cdot e_2)^3 ]+  c_{2,10}^{ (i j)} \  I_{i j}[\mu^2 ((q+p_i)\cdot e_2)]  \bigg\} \pagebreak[1]   \nn
&+&
\sum_{i}^{n-1} \bigg \{ c_{1,14}^{ (i)} \ I_{i}[\mu^2] 
+ c_{1,15}^{ (i)}  \    I_{i}[((q+p_i)\cdot e_3) ((q+p_i)\cdot e_4)   ]  \bigg\}
 \ .
\label{eq:AresultE}
\eea
The   integral  $I_{i j}[ ((q+p_i)\cdot e_2)^3]$ can be obtained from
the analytic expression of the rank-3 bubble given in section 4
of~\cite{Denner:2005nn}.  
The integrals 
$
I_{i}[\mu^2] ,  \  
I_{i}[((q+p_i)\cdot e_3) ((q+p_i)\cdot e_4)], \  
$
$
I_{i j}[\mu^2 ((q+p_i)\cdot e_2)],  
$
and
$
 I_{i j k}[\mu^4]
$ 
  are computed in  Appendix~\ref{App:IntEx}.

%% file: conclusions.tex
\section{Conclusions}

In this paper we presented a procedure for the semi-analytic reduction 
of one-loop scattering amplitudes in dimensional regularization, 
which exploits the asymptotic behavior of the integrand-decomposition.

The algorithm is based on a partial reconstruction of the numerator,
where the coefficients of the master integrals are determined
through a simplified integrand-reduction. Whenever necessary,
the complete integrand reconstruction can be achieved as well.

The analytic informations allow to  avoid the computation of   5-point coefficients 
and of the spurious 4-point ones. Moreover the  4-point non-spurious coefficients 
do not enter the determination of the lower-point ones. The integrand reduction
algorithm is indeed required only for the coefficients of 3-, 2-, and 1-point functions.

The  asymptotic  expansion makes both the  
numerator and the subtraction terms separately polynomial.
Therefore the subtraction of higher-point residues can be omitted during the reduction
and replaced by coefficient-level corrective terms.
The latter can be determined a priori  from the Laurent expansion of the 
expression of the  integrand-subtraction terms. Therefore the  actual reduction
algorithm applies only to the terms involving the 
numerator, 
whose reconstruction  is achieved  by {\it polynomial division}.
The coefficients of the 3-, 2-, and 1-point functions are finally obtained as trivial combinations 
of the coefficients coming out of the polynomial division and the corrective coefficients.

This method exploits as much as possible the known analytic structure of the integrand, hence 
it relies on the analytic structure of the numerator and 
its asymptotic expansions, used as input. It has been implemented and tested in  {\mathematica}
and  in {\C++}, using the polynomial division.
The semi-analytic implementation in {\C++} has been designed as a  reduction library to be linked
to codes like {\Gosam} and {\Formcalc} which generate analytic expressions for the integrands, as well as 
to any package providing the tensor-structure of the integrand.

We also discussed the extension of integrand-reduction methods to theories allowing for integrands 
with powers of the loop momentum larger then the number of denominators. 
We explicitly presented the extended polynomials in the case of powers larger
by one unit than the number of denominators. The advantages 
of the method hereby discussed are not spoiled by the presence of higher rank numerators.

We are confident that the investigation of the asymptotic regimes can
ameliorate the integrand decomposition of scattering amplitudes beyond one-loop as well.

%% file: AppendixN.tex
\newpage
\section{Numerical examples}
\label{App:Nume}

In this appendix we present two modest numerical applications of the hereby discussed algorithm. 
We compare the results with the ones obtained using the standard $d$-dimensional integrand 
reduction implemented in \samurai\ .
A comprehensive comparison between the two algorithms is beyond the scope of this paper.
We only intend to show the potential benefits arising from a {\it lighter} algorithm,
which requires less coefficients and which uses subtraction at the coefficient-level rather than
at the integrand-level.

\newcommand{\trule}{\rule[-1.5mm]{0mm}{6mm}}

\TABULAR[t]{llr}{
  \hline
   \hline\trule
  \multirow{2}{*}{}  &Result &   
\\
 \hline\trule
  &   contribution $a_{-2}$&\\
  &\samurai  & $-185.051790779988+ i \,2.1\times 10^{-12}$ \\
  & new algorithm    & $-185.051790779978- i \,9.0\times 10^{-14}$ \\ 
 \hline\trule
  &    contribution $a_{-1}$ &\\
  & \samurai  & $749.007288547- i \,580.971272485$  \\
  & new algorithm    & $749.007288566- i \,580.971272508$    \\
 \hline\trule
  & contribution  $a_0$ &\\
  & \samurai  &  $-724.020439522+ i \,2350.630383583$   \\
  & new algorithm    &  $-724.020439861 + i \,2350.630383784$   \\
 \hline
 \hline \trule
 &  coefficients evaluated &\\
&\samurai  & $461$  \\
& new algorithm    & $386$ \\ 
  \hline
 \hline 
}
{Integrand reduction of $\mathcal{I}_{6,6}$. The  comparison between \samurai\ and the new algorithm is shown. 
\label{Tab:Comp1}}

\subsection*{Rank six 6-point integrand}
We consider a six-point integrand of rank-6:
\begin{equation}
 \mathcal{I}_{6,6}= \frac{\mathcal{N}(q,\mu^2)}{ \prod_{i=0}^6 D_i} \, , \qquad  \mathcal{N}(q,\mu^2)= \prod_{i=1}^6 (q\cdot r_i)
\end{equation}
The  momenta appearing the denominator $D_i$ are
\small{
\bea
p^\mu_0 &=& ( 0,   0,   0,   0) \nn
p^\mu_1 &=&  (-56.6251094805816, 0,0,-56.6251094805816) \nn
p^\mu_2 &=&  (-113.250218961163, 0,0,0) \nn
p^\mu_3 &=&  (-68.5281885958052,  33.5,   15.9,   25) \nn
p^\mu_4 &=& (-48.7688869887140, 21, 31.2,   25.3) \nn
p^\mu_5 &=& (-27.9148705889889, 11, 13.2, 22) 
\eea
}
while all the masses are assumed to be vanishing. The momenta $r_i$  are given by
\small{
\bea
r_1^\mu &=& (1.30,5.10,0.50,0.40) \nn
r_2^\mu &=& (0.80,1.00,2.30,2.50) \nn
r_3^\mu &=& (1.90,3.20,1.77,2.11)  \nn
r_4^\mu &=& (3.03,1.05,2.33,1.77) \nn
r_5^\mu &=& (3.56,5.30,3.09,2.34) \nn
r_6^\mu &=& (7.08,1.98,5.30,4.55).
\eea
}
The integrated result is given as a series in $\epsilon = (4-d)/2$,
\bea
{(2 \pi \mu^2)^{(4-d)} \over i \pi^2} \int d^d {\bar q} \ {\cal I}_{6,6} = {a_{-2} \over \epsilon^2} + {a_{-1} \over \epsilon^1} + {a_0} + {\cal O}(\epsilon) \ .
\eea
In Table~\ref{Tab:Comp1} we show the numerical values of the coefficients $a_i$ computed with 
the algorithm implemented in \samurai\ and the new one. 
The two algorithms are in good agreement, but the new algorithm requires the determination of 386 
coefficients, instead of the 461 required by the standard reduction.

We estimate the quality of the reconstruction of the numerator using
the {\it global} $(N=N)-$test described in Section 3.4.1 of \cite{Mastrolia:2010nb}.
For this integrand, the reconstruction of the new algorithm is two digits more accurate than 
the one performed by \samurai.

\FIGURE[t]{
\includegraphics[width=7.4cm, height=6.3cm]{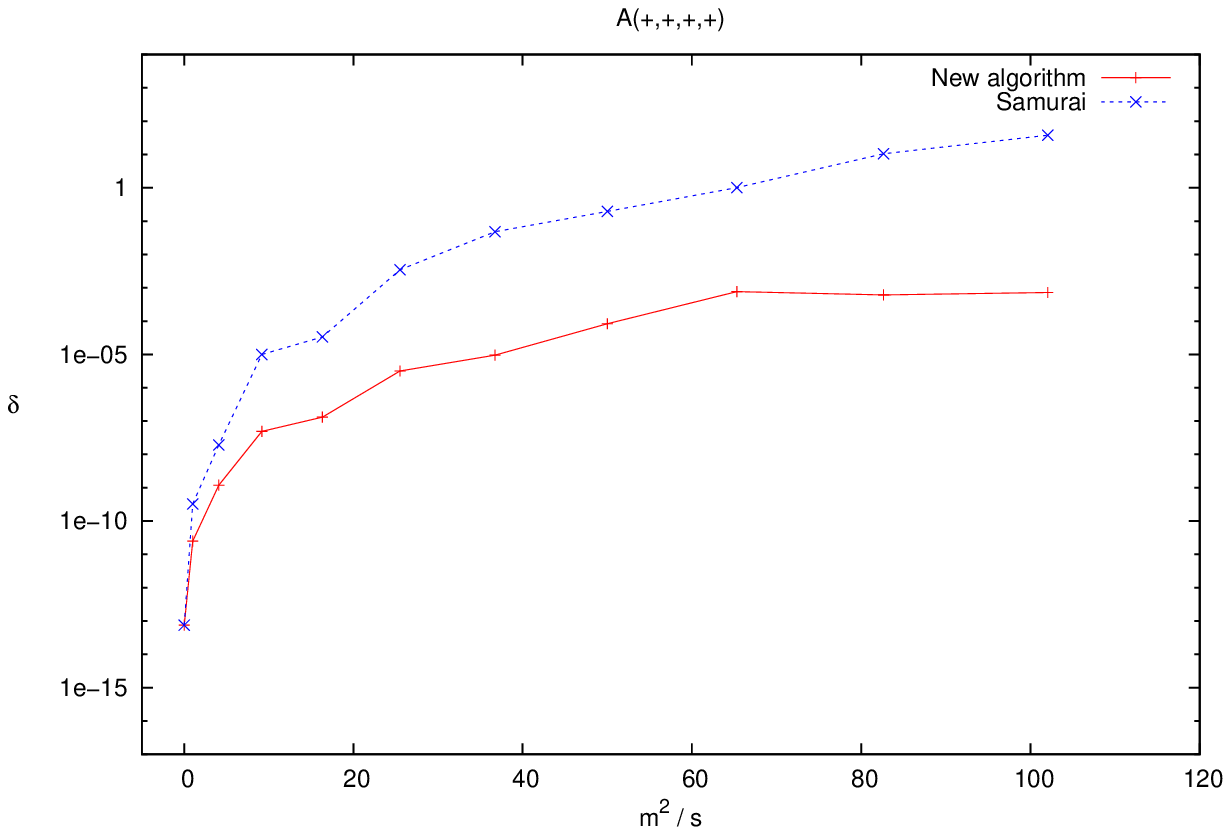}
\includegraphics[width=7.4cm, height=6.3cm]{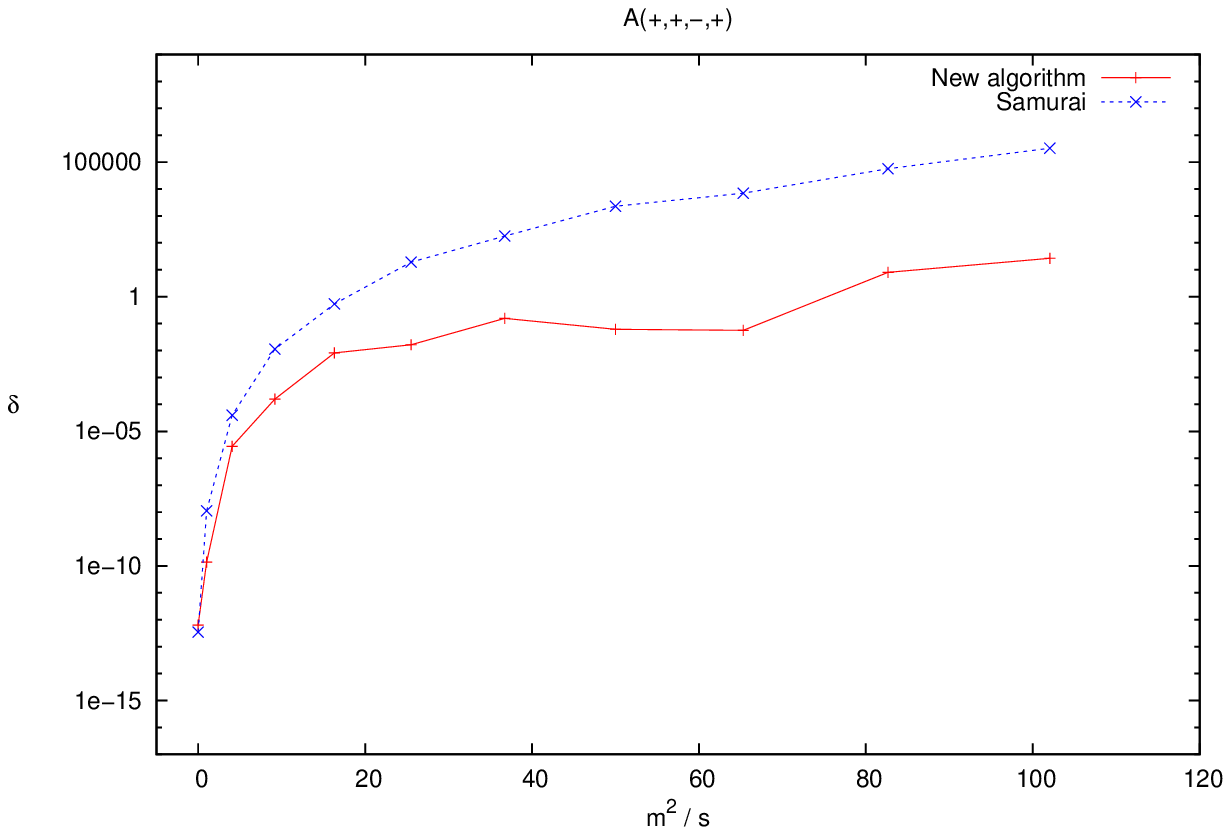}
\includegraphics[width=7.4cm, height=6.3cm]{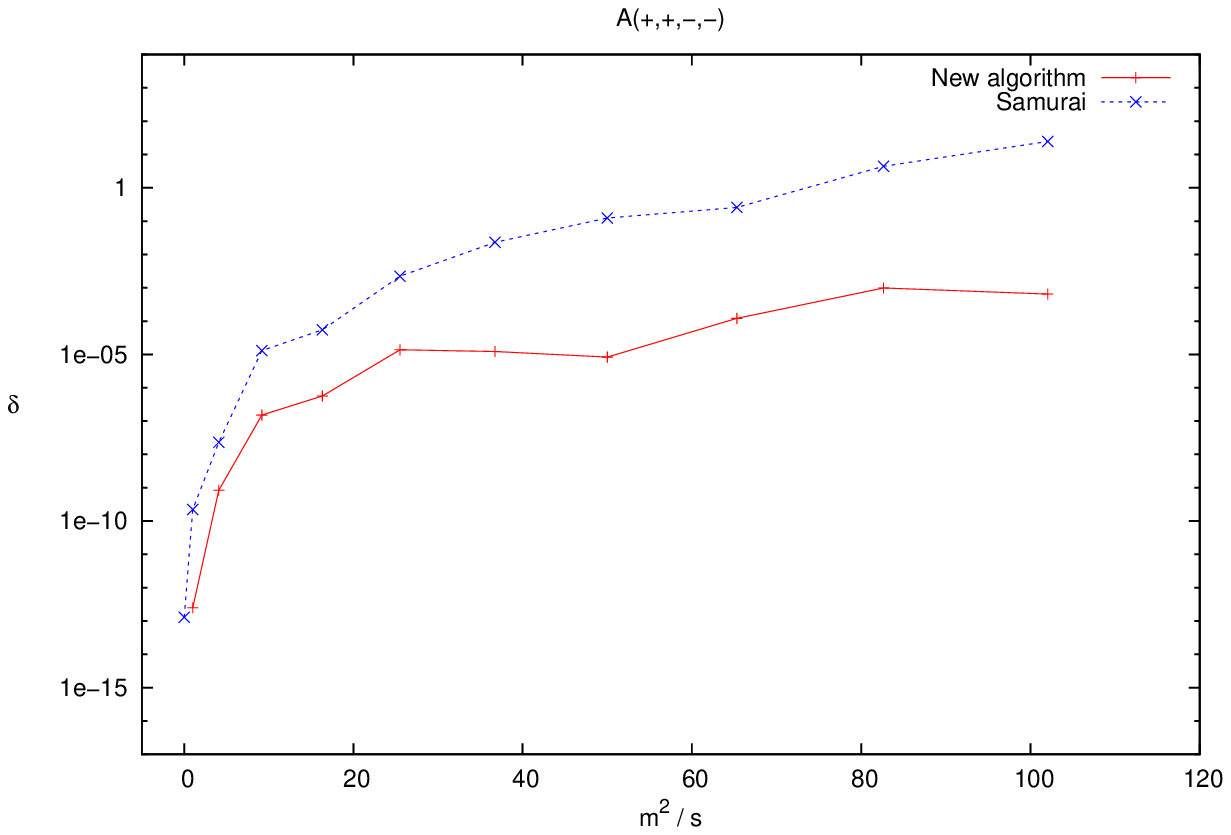}
  \caption{Leading order contribution to the amplitude $\mathcal{A}$.
  For each helicity configuration we plot the quantity $\delta$, eq.~(\ref{Eq:Delta4G}),
as a function of $m^2/ s$.   The numerical evaluation of $\mathcal{A}$ has been performed  using 
either  \samurai\ or  the new algorithm.
}
  \label{fig:A4G}
}

\subsection*{Fermion-mass dependence of the 4-photons amplitude}
Because of the presence of products of denominators in Eq.(\ref{def:OPP:deco}),
the numerical integrand  reconstruction may become unstable
if the internal masses are larger then the kinematic  invariants~\cite{Mastrolia:2010nb,Hirschi:2011pa}.
A simple example where such a situation may occur is the four-photon scattering in QED,
\bea
0 \;  \to \;   \gamma(k_1, h_1) \;  \gamma(k_2, h_2) \;  \gamma(k_3, h_3) \;  \gamma(k_4, h_4)  \, ,
\label{Eq:Pro4G}
\eea
where $k_i$ ($h_i$) denotes  the momentum (helicity) of the corresponding particle. 
We define $s \equiv (k_1+k_2)^2$.  
The leading-order process proceeds via fermionic-loop, and  
in the case of a single fermion of mass $m$, there are three 
independent helicity amplitudes 
\bea
\mathcal{A}(+,+,+,+), \qquad \mathcal{A}(+,+,-,+), \qquad \mathcal{A}(+,+,-,-),
\eea
which are known analytically~\cite{Gounaris:1999gh}. 
Mutual cancellations among contributions from different MI's render the numerical evaluation of these
amplitudes unstable, in particular when the ratio $m^2/s$ becomes large.  In this Appendix 
we explore the behaviour of the new algorithm in this kinematic regime, by comparing it to the one implemented 
in \samurai\ . We consider the phase space point
{\small
\bea
   k^\mu_1 &=&  (-7.0, 0.0, 0.0,-7.0) \nn
   k^\mu_2 &=& ( -7.0, 0.0, 0.0, 7.0) \nn
   k^\mu_3 &=& (  7.0, 6.1126608202785198  ,
-0.8284979592001092   ,   3.3089226083172685 ) \nn
  k^\mu_4 &=& ( 7 ,  -6.1126608202785278 ,
0.8284979592001093  ,   -3.3089226083172703 ) \ ,  
\eea
}
and we vary the numerical value of $m$.
In Figure~\ref{fig:A4G}
 we plot the relative difference $\delta$, defined as
 \bea
\delta \equiv  \left | \frac{\mathcal{A}_{\rm num} -\mathcal{A}_{\rm ana} }{\mathcal{A}_{\rm ana}}  \right | \,  ,
\label{Eq:Delta4G}
\eea
as a function of $m^2/ s$.
$\mathcal{A}_{\rm ana}$ ($\mathcal{A}_{\rm num}$)  is the analytical (numerical) value of the amplitude $\mathcal{A}$.
For each helicity configuration the new algorithm seems to be less affected by this kind of inaccuracy than the one currently implemented in \samurai\ .

%% file: AppendixM.tex
\section{Higher-rank  integrals}
\label{App:IntEx}
In this appendix we compute the higher-rank master integrals appearing eq.~(\ref{eq:AresultE}), 
\bea
 I_{i }[\mu^2] ,\qquad I_{i}[((q+p_i)\cdot e_3)  ((q+p_i)\cdot e_4)], \qquad  I_{i j}[\mu^2 ((q+p_i)\cdot e_2)] , \qquad  I_{ijk}[\mu^4] . 
 \label{Eq:IntApp}
\eea
The strategy is outlined in the Appendix I of~\cite{Bern:1995ix}. The 
integrals~(\ref{Eq:IntApp}) are obtained by projections, namely contracting appropriate 
tensors with the covariant decomposition of suitably chosen tensor integrals.

For later convenience we split the $d$-dimensional metric tensor $\hat g^{\mu \nu}$
into the $4$-dimensional part, $g^{\mu \nu}$ , an the $(-2\epsilon)$-dimensional part, $\tilde g^{\mu \nu}$.
We have 
\bea
\hat g^{\mu \nu} = g^{\mu \nu}  +  \tilde g^{\mu \nu}, \quad
\tilde g^{\mu \nu} \hat g_{\mu \nu} = -2 \epsilon, \quad 
\tilde g_{\mu \nu } q^{\mu} q^{\nu} = -\mu^2, \quad 
\tilde g_{\mu \nu } k^{\mu} =0, 
\eea
for each 4-dimensional vector $k^\mu$.

\subsection*{Integral $I_{i}[\mu^2]$}
The covariant decomposition of a rank-2 tadpole is
\bea
I^{\mu \nu}_{i } \equiv   \int d^d\bar q \frac{(\bar q +p_i)^\mu ( \bar  q+p_i)^\nu }{D_{i}} = \hat g^{\mu \nu} A_{00}
\label{Eq:CovTad2}
\eea
After the contraction with $\tilde g^{\mu \nu}$ we have
\bea
I_{i }[\mu^2]  =   \int d^d\bar q \frac{\mu^2}{D_i} =  2 \epsilon A_{00}  = \frac{ i \pi^2 m^4_i}{2}+\mathcal{O}(\epsilon).
\eea
The analytic expression of $A_{00}$ and  of the other tensor coefficients  appearing in this section can be found in~\cite{Denner:2005nn}.

\subsection*{Integral $I_{i}[((q+p_i)\cdot e_3)  ((q+p_i)\cdot e_4)]$}
The expression of this integral can be obtained  by contracting the covariant decomposition~(\ref{Eq:CovTad2}) with $e_3^{\mu} e_4^{\nu}$
and by using the relation $e_3\cdot e_4 =-1$. The outcome is
\bea
I_{i }[(q+p_i)\cdot e_3)  ((q+p_i)\cdot e_4)]  &=&   \int d^d\bar q \frac{(q+p_i)\cdot e_3)  ((q+p_i)\cdot e_4)}{D_i}  = 
(e_3\cdot e_4) A_{00} \nn
&=& -   \frac{m_i^2 \, I_i + I_i[\mu^2]}{4}  +\mathcal{O}(\epsilon) \, .
\eea

\subsection*{Integral $I_{ij}[\mu^2((q+p_i)\cdot e_2)]$} 
The covariant decomposition of a rank-3 bubble reads as follows
\bea
I^{\mu \nu \rho}_{ij } \equiv   \int d^d\bar q \frac{(\bar q +p_i)^\mu (\bar q+p_i)^\nu (\bar
q+p_i)^\rho}{D_{i}D_{j}} &=& \Big( \hat g^{\mu\nu} \, p_{ji}^\rho + \hat g^{\mu\rho}\,  p_{ji}^\nu  +\hat  g^{\nu\rho}\,  p_{ji}^\mu \Big) B_{001} \nn 
&+&  p_{ji}^\mu  p_{ji}^\nu  p_{ji}^\rho\, B_{111}, 
\eea
with $p_{ji} \equiv p_j-p_i$. After the contraction with $\tilde g^{\mu \nu} e_2^\rho$ we have
\bea
  I_{i j}[\mu^2\, ((q+p_i)\cdot e_2)] &=&  \int d^d\bar q \frac{\mu^2\,
((q+ p_i)\cdot e_2)}{D_{i}D_{j}} = 2\epsilon\, (p_{ji} \cdot e_2)\, B_{001} \nn
&=& \frac{i \pi^2}{12} \, (p_{ji} \cdot e_2)\,  \left (  p_{ji}^2 -2 m^2_i -4m^2_j \right )+\mathcal{O}(\epsilon) \, .
\eea

\subsection*{Integral $I_{ijk}[\mu^4]$} 
We start from the decomposition
\bea
I^{\mu \nu \rho \sigma}_{ijk } \equiv   \int d^d\bar q \frac{(\bar q+p_i)^\mu (\bar q+p_i)^\nu( \bar q+p_i)^\rho (\bar
q+p_i)^\sigma}{D_{i}D_{j}D_{k}} & =& \Big( \hat g^{\mu\nu}  \hat g^{\rho\sigma} +
\hat g^{\mu\rho} \hat g^{\nu\sigma} +\hat  g^{\mu\sigma} \hat g^{\nu\rho} \Big) C_{0000} \nn
&+& \left ( \mbox{rank-4 tensors 
containing } p_{ji} \, ,  p_{jk}  \right ) \quad
%
\eea
After the contraction with $\tilde g^{\mu\nu}\,
\tilde g^{\rho\sigma}$ the tensors containing 
$p_{ji}, \, p_{jk}$ vanishes and we get 
\bea
  I_{ijk}[\mu^4] &=& \int d^d\bar q
\frac{\mu^4}{D_{i}D_{j}D_{k}} = 4\epsilon(\epsilon-1)\, C_{0000} \nn
&=& \frac{i \pi^2}{6} \left [
\frac{p^2_{jk} + p^2_{ji} + p^2_{ki}   }{4} - m^2_i - m^2_j  - m^2_k
\right ] + \mathcal{O}(\epsilon) \, .
\eea